\documentclass[lettersize,journal]{IEEEtran}
\usepackage{amsmath,amsfonts}
\usepackage{mathrsfs}
\usepackage[ruled,linesnumbered]{algorithm2e}
\usepackage{array}
\usepackage[caption=false,font=normalsize,labelfont=sf,textfont=sf]{subfig}
\usepackage{textcomp}
\usepackage{stfloats}
\usepackage{url}
\usepackage{verbatim}
\usepackage{graphicx}
\usepackage{cite}
\usepackage{amssymb}
\usepackage{cuted}
\usepackage{color}
\usepackage{eurosym,booktabs,multirow}
\usepackage{verbatim}
\usepackage{cite}
\usepackage{makecell}
\usepackage{subfig}
\usepackage{stfloats}
\usepackage{enumitem}

\begin{document}

\title{Analysis on Energy Efficiency of RIS-Assisted Multiuser Downlink Near-Field Communications}

\author{Wei~Wang,~\IEEEmembership{Graduate Student Member,~IEEE}, Xiaoyu~Ou,~\IEEEmembership{Student Member,~IEEE}, Zhihan~Ren,~\IEEEmembership{Graduate Student Member,~IEEE}, Waqas~Bin~Abbas, Shuping~Dang,~\IEEEmembership{Senior Member,~IEEE}, Angela~Doufexi,~\IEEEmembership{Member,~IEEE}, and Mark~A.~Beach,~\IEEEmembership{Senior Member,~IEEE}
\thanks{This work was supported by China Scholarship Council (CSC) under grant 202108060224.}
\thanks{The authors are with the Communication Systems and Networks Group, School of Electrical, Electronic and Mechanical Engineering, University of Bristol, Bristol BS8 1UB, U.K. (e-mail:~wei.wang@bristol.ac.uk; xiaoyu.ou@bristol.ac.uk; z.ren@bristol.ac.uk; waqas.abbas@bristol.ac.uk; shuping.dang@bristol.ac.uk; a.doufexi@bristol.ac.uk; and m.a.beach@bristol.ac.uk).}
}



\maketitle

\begin{abstract}
In this paper, we focus on the energy efficiency (EE) optimization and analysis of reconfigurable intelligent surface (RIS)-assisted multiuser downlink near-field communications. Specifically, we conduct a comprehensive study on several key factors affecting EE performance, including the number of RIS elements, the types of reconfigurable elements, reconfiguration resolutions, and the maximum transmit power. To accurately capture the power characteristics of RISs, we adopt more practical power consumption models for three commonly used reconfigurable elements in RISs: PIN diodes, varactor diodes, and radio frequency (RF) switches. These different elements may result in RIS systems exhibiting significantly different energy efficiencies (EEs), even when their spectral efficiencies (SEs) are similar. Considering discrete phases implemented at most RISs in practice, which makes their optimization NP-hard, we develop a nested alternating optimization framework to maximize EE, consisting of an outer integer-based optimization for discrete RIS phase reconfigurations and a nested non-convex optimization for continuous transmit power allocation within each iteration. Extensive comparisons with multiple benchmark schemes validate the effectiveness and efficiency of the proposed framework. Furthermore, based on the proposed optimization method, we analyze the EE performance of RISs across different key factors and identify the optimal RIS architecture yielding the highest EE.
\end{abstract}

\begin{IEEEkeywords}
Energy efficiency (EE), reconfigurable intelligent surface (RIS), near-field communications, multiuser communications, discrete phase, alternating maximization.
\end{IEEEkeywords}

\section{Introduction}
\IEEEPARstart{R}{econfigurable} intelligent surface (RIS) is one of the most promising enabling technologies for future sixth-generation (6G) wireless communication networks~\cite{9140329}, due to its outstanding capabilities for various applications, including filling dead zones, extending coverage, improving channel conditions, providing customized quality of service (QoS), and ensuring secure transmission.An RIS is composed of a large number of passive elements that can reflect or refract impinging signals in desired directions, enabling the adaptability of wireless propagation environment~\cite{9424177}.
The passive nature of elements results in RIS-assisted communications with low fabrication costs and operating energy consumption because they do not require the radio frequency (RF) chain and power amplifiers.Since current fifth-generation (5G) networks have exhibited higher power consumption than their predecessors~\cite{10207809,10247147}, RISs have also emerged as an energy-efficient paradigm in the evolution of mobile technology to align with the Net-Zero goal, which aims to reduce global carbon emissions by half by 2030 and achieve net zero emissions by 2050~\cite{10247147}.
These aforementioned features have attracted considerable attention towards exploring the potential of RIS-assisted communications in recent years, with a primary focus on far-field application scenarios~\cite{zhi2022two, wang2021joint, huang2019reconfigurable, long2021active, you2020energy, pei2021ris, tang2022path, wang2024reconfigurable}.

It is anticipated that extremely large-scale RISs (XL-RISs) are expected to play a crucial role in future 6G communications by better compensating for severe path loss and ensuring capacity improvement~\cite{9810144}, which shall encompass near-field communication applications.
With the increase in aperture length of XL-RIS as well as the operating frequency, the Fraunhofer distance~\cite{selvan2017fraunhofer}, which distinguishes the near-field region and the far-field region, also extends, raising up the probability that the transmitters and receivers are in the near-field region of RIS-assisted communications. In this case, far-field channel models become inapplicable because the electromagnetic wavefronts can no longer be approximated as planar waves, which is a fundamental assumption of far-field analysis. For characterizing near-field communications, precise channel models must be able to capture the electromagnetic effects of RF signals over the three-dimensional coordinates of each transmitter antenna, receiver antenna, and element of RIS.
Compared with far-field communications, near-field communications have the advantage of providing high-rank line-of-sight (LoS) channels for increased degrees of freedom and concentrating the signal's energy on the target receivers to prevent leakage~\cite{10380596}. 
Currently, the near-field communication capabilities of RIS have been explored in localization~\cite{10113892, 9650561, 9709801}, channel estimation~\cite{10500431, 10081022, 10462912}, beam management~\cite{shen2023multi, 10130629, 9941256}, and achievable rate improvement~\cite{10221794}.

To the best of our knowledge, however, there has been little study on the energy efficiency (EE) issues of RIS-assisted near-field communications. Additionally, current studies on far-field energy-efficient RIS-assisted communications, including~\cite{hou2020reconfigurable,9174801,yang2021energy,huang2019reconfigurable,you2020energy,zhang2021beyond}, oversimplify their power consumption models, failing to reveal the practical power consumption characteristics of RIS, and therefore, giving only limited insights into the EE of these novel communication paradigms. 
For instance, \cite{hou2020reconfigurable,9174801,yang2021energy} related the power consumption of the RIS boards solely to the number of reconfigurable elements, and \cite{huang2019reconfigurable,you2020energy,zhang2021beyond} further considered the impact of RIS reconfiguration resolutions. However, apart from the element number and resolution, the power dissipated by RIS boards is also distinguished by different methods for realizing reconfigurability. This is because reconfigurable elements have different power consumption characteristics due to their working mechanisms.
For example, varactor-diodes are operated by energy-greedy driving circuits, but the power dissipation on the elements themselves is negligible~\cite{pei2021ris,tang2022path}. In contrast, the power consumption of PIN diodes and RF switches is significantly lower by their driving circuits but non-negligible by adjusting elements~\cite{wang2024reconfigurable,zhang2019breaking,rossanese2022designing,rao2022novel}.
Furthermore, though both PIN diodes and 1-bit RF switches have two operating states that can be expressed as `0' and `1', the power dissipation of PIN diodes increases with the number of PIN diodes operating in state `1', whereas 1-bit RF switch-based reconfigurable elements has similar power dissipation no mater in which operating state~\cite{wang2024reconfigurable}.

Since passive RISs only reflect or refract impinging signals without processing, \cite{bjornson2019intelligent} has revealed that RISs must exceed a certain number of elements to outperform relaying. While the power consumption differences may be negligible for individual RIS elements, they become significant in large-scale RIS arrays, especially in near-field scenarios and XL-RISs envisioned for future 6G networks~\cite{9810144,10380596}. As RIS size grows, its power consumption becomes a critical factor in overall system efficiency. In this context, even if RIS arrays using different reconfiguration methods achieve the same spectral efficiency under identical configurations, their system energy efficiencies can vary significantly due to distinct switch power characteristics, which cannot be captured by the simplified models in prior literature.
Moreover, the power dissipation of PIN diodes depends on real-time RIS configurations, leading to a coupling between system power consumption and instantaneous RIS configurations. Consequently, both spectral efficiency (SE) and power consumption become functions of the RIS real-time configuration, the discrete phase shift nature of which further increases the optimization complexity. Meanwhile, RIS configuration optimization in EE analysis must balance SE maximization and power consumption minimization, rather than solely focusing on SE maximization.
These factors highlight the need for a comprehensive optimization and analysis of all possible factors affecting RIS power consumption to bridge the gap between theoretical EE analysis and real-world performance, enabling a more precise assessment of RIS design trade-offs for energy-efficient wireless systems.

\subsection{Contributions}
In this paper, we study the EE of RIS-assisted near-field wireless communication systems. We adopt a more practical power consumption model modified from the hardware measurements in~\cite{wang2024reconfigurable, pei2021ris}, optimize, and analyze the system's EE under various reconfigurable element numbers, resolutions, reconfiguration methods, and maximum transmit power. The main contributions of this work are summarized as follows:
\begin{itemize}[leftmargin=*]
\item
We construct an RIS-assisted multiuser downlink near-field communication system, in which a more practical power consumption model for element-wise independently controlled RISs is introduced. This model considers a comprehensive set of factors, including the number of RIS elements, the reconfigurable resolution of each RIS element, realistic power dissipation characteristics of multiple reconfiguration switches (i.e., PIN diodes, varactor diodes, and RF switches) in the reconfiguration and driving circuits, the RIS controller, and the base station (BS) power allocation. To better quantify and evaluate different reconfiguration switches, we model the power consumption of a multi-bit RF switch as an assembly of multiple 1-bit switches within each RIS element, allowing a fair comparison of the power dissipation characteristics of RIS elements across various reconfiguration methods and resolutions. To the best of our knowledge, this work is the first to provide a quantitative, side-by-side evaluation of the EE of RIS-assisted networks using such a practical power consumption model for different reconfiguration methods.

\item
Considering the practical limitation that reconfigurable elements cannot be adjusted continuously but instead operate in finite discrete states depending on the reconfiguration resolution, as well as the coupling between RIS configurations and RIS power consumption due to the power dissipation characteristics of PIN diodes, we propose a novel low complexity nested alternating maximization algorithm framework to achieve optimal system EE. In this framework, the RIS array configuration is optimized using integer particle swarm optimization (PSO), where each particle is an integer matrix that directly explores potential RIS configurations within a discrete parameter space. Meanwhile, a power allocation scheme is designed to optimize the BS transmit power over a continuous parameter space using a non-convex algorithm, which is nested within the iterative process of the integer PSO. Compared with conventional continuous-phase optimization, the proposed discrete-phase-optimized RIS demonstrates more consistent performance in practical communication systems.

\item
We numerically evaluate the performance of the proposed algorithm and investigate multiple key factors affecting the EE of RIS-assisted systems in a near-field indoor scenario. Comparisons with benchmark schemes validate the effectiveness and efficiency of the proposed algorithm in the discrete phase-shift optimization. Furthermore, based on the simulation results, we provide a comprehensive analysis of the EE performance across various RIS array sizes, reconfiguration methods, resolutions, and maximum transmit power. Additionally, we examine the trade-off between system EE and RIS complexity, identifying the optimal RIS architecture setups for different scenarios.
\end{itemize}
\vspace{-10pt}

\subsection{Organization}
In the remainder of this paper, the signal model, near-field channel model, precoding, and power consumption model are given in Section~\ref{sec:system model}. Then, the optimal EE of the RIS-assisted near-field communication system is formulated in Section~\ref{sec:problem formulations and solutions}, solved by the proposed alternating optimization framework, which includes discrete optimization for RIS configuration and continuous optimization for BS power allocation. In Section~\ref{sec:numerical results and snalysis}, we detail the simulation setups and present numerical results to evaluate and verify the effectiveness of the proposed optimization approach. Furthermore, we provide an in-depth analysis of the key factors affecting the EE of the RIS-assisted systems. Finally, Section~\ref{sec:conclusions} concludes this paper.
\vspace{-5pt}

\section{System Model}
\label{sec:system model}
\subsection{Signal Model}
\label{subsec:signal model}
In near-field communication networks, the communication distances are relatively short, and BSs, RISs, and user equipment (UE) are placed closely together in confined indoor or outdoor spaces. Given these conditions, femtocell BSs (FBSs) are adopted as the transmitters in this paper due to their compact size, low power consumption, and cost-effective deployment. Typically, these FBSs can provide reliable communication services within coverage radii ranging from 20 to 30 meters~\cite{9082809,9246508}.

Under the assumption that the direct paths from FBS to UEs are blocked because of the abundant objects in complex indoor environments, we consider an RIS-assisted indoor wireless communication scenario with $M$-element FBS and $N$-element RIS, both cooperating to serve $K$ single-antenna UEs. For the $k$th UE with $k \in K$, received signal $y_k$ can be expressed as
\begin{equation}
    y_{k} = (\boldsymbol{\rm{h}}_{r,k}^{T} \boldsymbol{\Phi}\it{\boldsymbol{\rm{G}}}) \boldsymbol{\rm x} + \it{n}_{k},
\label{equi:signal_1}
\end{equation}
where $\boldsymbol{\rm{h}}_{r,k} \in \mathbb{C}^{N}$ and $\boldsymbol{\rm{G}} \in \mathbb{C}^{N\times M}$ denote the reflected channel vector between the $k$th UE and the RIS and the incident channel matrix between the FBS and the RIS, respectively; $\boldsymbol{\Phi} \in \mathbb{C}^{N\times N}$ indicates the beamforming matrix of the RIS, and $\it{n}_{k} \in \mathbb{C}$ is the received noise with zero mean and variance $\sigma^{2}_{k}$ at $k$th UE;
$\boldsymbol{\rm{x}} \in \mathbb{C}^{M}$ is the precoded transmitted signal vector at the FBS which can be expressed as
\begin{equation}
    \boldsymbol{\rm{x}} = \sum_{k=1}^{K} \sqrt{P_k} \boldsymbol{{\rm \omega}}_{k}s_{k},
\label{equi:precoded}
\end{equation}
where $s_{k} \in \mathbb{C}$ denotes the desired signal to the $k$th UE; $\boldsymbol{\rm{\omega}}_{k} \in \mathbb{C}^{M}$ is its precoding vector, and $P_k$ is the transmit power.
Beamforming matrix $\boldsymbol{\Phi}$ at the RIS is a diagonal matrix which can be expanded as
\begin{equation}
    \boldsymbol{\Phi} \triangleq {\rm diag}(\boldsymbol{\phi})= \rm{diag}({[\beta_{1}e^{j\theta_{1}},\beta_{2}e^{j\theta_{2}},...,\beta_{N}e^{j\theta_{N}}]}^{\it{T}}),
\label{equi:RIS}
\end{equation}
where $\beta_{n}$ is the amplitude response, and $\theta_{n}$ is the phase shift of the $n$th reconfigurable element.

By combining  (\ref{equi:signal_1}) and (\ref{equi:RIS}), received signal $y_k$ can also be rewritten as
\begin{equation}
    y_{k} = (\boldsymbol{\phi}^{T} {\rm diag}(\boldsymbol{\rm{h}}_{r,k}) {\boldsymbol{\rm{G}}}) \boldsymbol{\rm{x}} + \it{n}_{k},
\label{equi:signal_2}
\end{equation}
where the reflecting channel is characterized by matrix $\boldsymbol{\rm{H}}_{k}\triangleq {\rm diag}(\boldsymbol{\rm{h}}_{r,k}) {\boldsymbol{\rm{G}}}$, and the cascaded channel when considering the RIS beamforming matrix can thus be characterized by vector $\boldsymbol{\rm{h}}_{\rm{ris}\it{,k}}\triangleq \boldsymbol{\phi}^{T} {\rm diag}(\boldsymbol{\rm{h}}_{r,k}) {\boldsymbol{\rm{G}}}$ for the $k$th UE.

Apart from blocking the direct path, the abundant objects and obstacles in our assumed environment will also bring about multi-paths in the FBS-RIS and RIS-UEs channels. Hence, similar to \cite{han2019large,zhi2022two,wang2021joint}, we adopt the Rician fading model to further capture the multi-path effects of the FBS-RIS and UE-RIS sub-channels with a dominant LoS path as
\begin{equation}
\boldsymbol{\rm{h}}_{r,k}=\sqrt{\frac{\varepsilon_{h}}{\varepsilon_{h}+1}}\overline{\boldsymbol{\rm{h}}}_{r,k}+\sqrt{\frac{1}{\varepsilon_{h}+1}}\widetilde{\boldsymbol{\rm{h}}}_{r,k},
\label{equi:Rician_1}
\end{equation}
and
\begin{equation}
\boldsymbol{\rm{G}}=\sqrt{\frac{\varepsilon_{G}}{\varepsilon_{G}+1}}\overline{\boldsymbol{\rm{G}}}+\sqrt{\frac{1}{\varepsilon_{G}+1}}\widetilde{\boldsymbol{\rm{G}}},
\label{equi:Rician_2}
\end{equation}
where $\varepsilon_{h}$ and $\varepsilon_{G}$ are the Rician factors that describe the weights of the LoS component in the propagation. Furthermore, $\overline{\boldsymbol{\rm{h}}}_{r,k} \in \mathbb{C}^{N}$ and $\overline{\boldsymbol{\rm{G}}} \in \mathbb{C}^{N\times M}$ represent the LoS components which are modeled depending on the spatial distance, whereas $\widetilde{\boldsymbol{\rm{h}}}_{r,k}$ and $\widetilde{\boldsymbol{\rm{G}}}$ signify the NLoS components which are assumed to be independently and identically distributed (i.i.d.).
\vspace{-10pt}

\subsection{Near-Field Channel Model}
\begin{figure}[t]
    \centering
    \includegraphics[width=0.7\columnwidth]{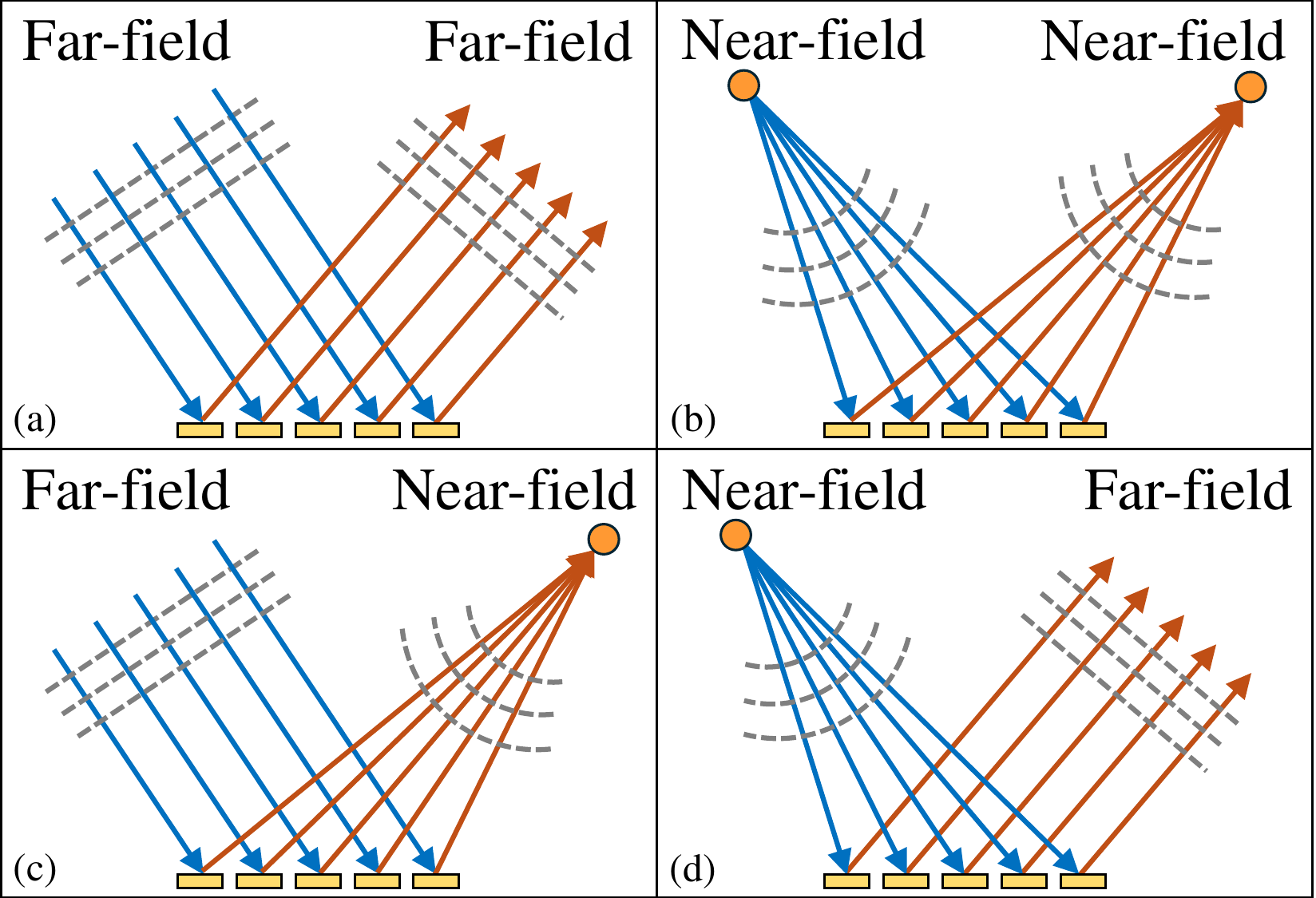}
    \caption{Four different scenarios of RIS, both the incidence and reflection are in the (a) far field or (b) near field, or one is in the far field while the other is in the near field in (c) and (d).}
    \label{fig:FF_and_NF}
\vspace{-10pt}
\end{figure}
The electromagnetic wave transmitted from an isotropic source displays a spherical wavefront. However, when the receiver is far away from the transmitter, the received wavefront can be approximated as planar because of the negligible curvature of the waves at the observation distance. The Fraunhofer distance distinguishes the spherical wavefront in the near-field Fresnel region and the plane wavefront in the far-field Fraunhofer region has been derived in \cite{selvan2017fraunhofer} as
\begin{equation}
    R_{\rm{F}} = \frac{2D^2}{\lambda},
\label{equi:Fraunhofer}
\end{equation}
where $D$ is the aperture length of the receiver, and $\lambda$ is the wavelength at the working frequency.
In the RIS-assisted cascaded channel, both the propagation between FBS and RIS and the propagation between RIS and UEs are considered, as they have four different situations as shown in Fig.~\ref{fig:FF_and_NF}. Defining these two propagation distances as $R_1$ for the FBS-RIS channel and $R_2$ for the RIS-UE channel, respectively, the near-field criterion from (\ref{equi:Fraunhofer}) can be rewritten according to \cite{cui2022near} as
\begin{equation}
    R_{\rm{F,RIS}} = \frac{2D_{\rm{RIS}}^2}{\lambda}\geq \frac{R_1R_2}{R_1+R_2},
\label{equi:Fraunhofer_RIS}
\end{equation}
where $R_{\rm{F,RIS}}$ is the Fraunhofer distance over the RIS-assisted propagation path, and $D_{\rm{RIS}}$ is the aperture length of the corresponding RIS.

With a pair of $R_1$ and $R_2$ meeting the criterion presented in (\ref{equi:Fraunhofer_RIS}), the curvature of the electromagnetic waves cannot be neglected and the propagation wavefront cannot be approximated as a planer as what has been normally done in the far-field propagation. In near-field propagation, the angles of departure (AoD) and the angles of arrival (AoA) are different for the links connecting each FBS antenna, reconfigurable element, and UE. Therefore, based on the three-dimensional coordinates of FBS antennas and reconfigurable elements, each vector $\overline{\boldsymbol{\rm{g}}}_{m} \in \mathbb{C}^{N}$ in the LoS component of downlink channel matrix $\overline{\boldsymbol{\rm{G}}}=[\overline{\boldsymbol{\rm{g}}}_{1},\overline{\boldsymbol{\rm{g}}}_{2},...,\overline{\boldsymbol{\rm{g}}}_{M}]$ between FBS and RIS in the near field can be constructed as
\begin{equation}
\overline{\boldsymbol{\rm{g}}}_{m} = \frac{1}{\sqrt{\boldsymbol{\alpha}_{\rm{in,m}}}} \odot \boldsymbol{\rm{a}}_{\rm{in,m}}(\boldsymbol{\rm{\varphi}}^{\rm{in}}_{n,m}),
\label{equi:BS_RIS}
\end{equation}
where $\odot$ is the Hadamard product, and $\boldsymbol{\alpha}_{\rm{in,m}} \in \mathbb{C}^{N}$ is the path-loss coefficients between the $m$th FBS antenna and RIS with the adoption of the ITU LoS basic transmission loss model in \cite{ITU}. Simultaneously, with $\mathcal{I}(N)=\{1,...,N\}$ representing the integer sequence, ${[\boldsymbol{\rm{a}}_{\rm{in}\it{,m}}(\boldsymbol{\rm{\varphi}}^{\rm{in}}_{n,m})]}_{n \in \mathcal{I}(N)}$ denotes the incident channel response vector from the $m$th FBS antenna to all reconfigurable elements and can be expanded as
\begin{equation}
\boldsymbol{\rm{a}}_{\rm{in}\it{,m}}(\boldsymbol{\rm{\varphi}}^{\rm{in}}_{n,m})=[e^{-j\varphi^{\rm{in}}_{1,m}},e^{-j\varphi^{\rm{in}}_{2,m}}, ...,e^{-j\varphi^{\rm{in}}_{N,m}}]^{T},
\label{equi:vector_in}
\end{equation}
where $\varphi^{\rm{in}}_{n,m}$ is the phase change from $m$th FBS antenna to the $n$th reconfigurable element, related to the space coordinate distance between these two components.

Accordingly, the LoS component of the downlink channel between the RIS and the $k$th UE in near field can be expressed as
\begin{equation}
\overline{\boldsymbol{\rm{h}}}_{r,k}^{T} = \frac{1}{\sqrt{\boldsymbol{\alpha}_{\rm{re}\it{,k}}}} \odot \boldsymbol{\rm{a}}_{\rm{re}\it{,k}}(\boldsymbol{\rm{\varphi}}^{\rm{re}}_{k,n}),
\label{equi:RIS_UE}
\end{equation}
where ${[\boldsymbol{\rm{a}}_{\rm{re}\it{,k}}(\boldsymbol{\rm{\varphi}}^{\rm{re}}_{k,n})]}_{n \in \mathcal{I}(N)}$ is given by
\begin{equation}
\boldsymbol{\rm{a}}_{\rm{re}\it{,k}}(\boldsymbol{\rm{\varphi}}^{\rm{re}}_{k,n})=[e^{-j\varphi^{\rm{re}}_{k,1}},e^{-j\varphi^{\rm{re}}_{k,2}},...,e^{-j\varphi^{\rm{re}}_{k,N}}].
\label{equi:vector_re}
\end{equation}

Different from calculating the phase shift differences of arrival based on the transmitter array steering vector and receiver array steering vector in the far-field scenario~\cite{shen2023multi}, we construct the near-field channel response with the help of the space coordinate distances between the transmitter and receiver. Therefore, the phase responses in (\ref{equi:vector_in}) and (\ref{equi:vector_re}) can be determined as
\begin{equation}
\varphi^{X}_{a,b} = \frac{2\pi}{\lambda} D^{X}_{a,b}\,,
\label{equi:phase_shift}
\end{equation}
where $\frac{2\pi}{\lambda}$ is the wave number, and $D^{X}_{a,b}$ is the space coordinate distance between $b$th element of transmitter and $a$th element of the receiver, which is explicitly given in (\ref{equi:OPD}).
In (\ref{equi:OPD}), $\boldsymbol{\rm{C}}^{Y_1}_{a} \in \mathbb{R}^{3}$ is the column vector recording the x-, y-, and z-coordinates of $a$th element in $Y_{[\cdot]}$, and ${[\cdot]}_{x}$ is the value of the vector subtraction at the x-coordinate. In the above equations, $X \in \{\text{in}, \text{re}\}$ indicates the incident channels and the reflected channels; $Y_1, Y_2 \in \{\text{FBS}, \text{RIS}, \text{UE}\}$ indicates the FBS, RIS, and multiple UEs, respectively.

\begin{figure*}[ht]
\begin{equation}
D^{X}_{a,b}=\sqrt{{[\boldsymbol{\rm{C}}^{Y_1}_{a}-\boldsymbol{\rm{C}}^{Y_2}_{b}]_{x}}^{2}+{[\boldsymbol{\rm{C}}^{Y_1}_{a}-\boldsymbol{\rm{C}}^{Y_2}_{b}]_{y}}^{2}+{[\boldsymbol{\rm{C}}^{Y_1}_{a}-\boldsymbol{\rm{C}}^{Y_2}_{b}]_{z}}^{2}}\,.
\label{equi:OPD}
\end{equation}
\hrulefill
\vspace{-10pt}
\end{figure*}

\subsection{Discrete RIS Configuration and Precoding}
To reduce the complexity of RIS architecture in both theoretical analysis and practical fabrication, here, we fix the amplitude and allow only the phase of reconfigurable elements to be adjusted for producing optimal performance. Thus, $\boldsymbol{\phi}$ in (\ref{equi:RIS}) can be rewritten as
\begin{equation}
    \boldsymbol{\phi}= \beta{[e^{j\theta_{1}},e^{j\theta_{2}},...,e^{j\theta_{N}}]}^{\it{T}} \triangleq \beta{[e^{j{\boldsymbol \theta}^{\rm cont}}]}^{\it{T}},
\label{equi:RIS_cont}
\end{equation}
where ${\boldsymbol \theta}^{\rm cont} = [\theta_{1}, \theta_{2}, ..., \theta_{N}]$ represents the continuous phase response applied to RIS for the RIS-assisted transmission enhancement. To analyze the system performance under discrete phase configurations of RIS, ${\boldsymbol \theta}^{\rm cont}$ in (\ref{equi:RIS_cont}) is converted to a discrete arrangement ${\boldsymbol{\theta}^{\rm{disc}}} \in \mathbb{R}^{1\times N}$ using quantization based on the RIS resolution. Meanwhile, $\boldsymbol{\rm{h}}_{\rm{ris}\it{,k}} \in \mathbb{C}^{1\times M}$ given in Section~\ref{subsec:signal model} can be rewritten as
\begin{equation}
\boldsymbol{\rm{h}}_{\rm{ris}\it{,k}}\triangleq \beta e^{j{\boldsymbol{\theta}^{\rm{disc}}}} {\rm diag}(\boldsymbol{\rm{h}}_{r,k}) {\boldsymbol{\rm{G}}}.
\label{equi:H_RIS_k}
\end{equation}
Cascaded channel matrix $\boldsymbol{\rm{H}}_{\rm{ris}} \in \mathbb{C}^{K\times M}$ with the RIS beamforming matrix for all \textit{K} UEs can be expressed as
\begin{equation}
\boldsymbol{\rm{H}}_{\rm{ris}} = {[\boldsymbol{\rm{h}}_{{\rm ris},1},\boldsymbol{\rm{h}}_{{\rm ris},2},...,\boldsymbol{\rm{h}}_{{\rm ris},K}]}^{T}.
\label{equi:H_RIS}
\end{equation}
After incorporating $\boldsymbol{\rm{H}}_{\rm{ris}}$ with ${\boldsymbol{\theta}^{\rm{disc}}}$, the regularized zero-forcing (RZF) is adopted to generate FBS precoding vector $\boldsymbol{\rm{\omega}}_{k}$ in (\ref{equi:precoded}), and combined matrix $\boldsymbol{\rm{\Omega}} = [\boldsymbol{\rm{\omega}}_{1},\boldsymbol{\rm{\omega}}_{2},...,\boldsymbol{\rm{\omega}}_{K}] \in \mathbb{C}^{M\times K}$ is given as
\begin{equation}
\boldsymbol{\rm{\Omega}}=\left\{
\begin{array}{cl}
\dfrac{\boldsymbol{\rm{H}}_{\rm{ris}}^{H}{(\boldsymbol{\rm{H}}_{\rm{ris}} \boldsymbol{\rm{H}}_{\rm{ris}}^{H} + \kappa \boldsymbol{\rm I}_{K})}^{-1}}{\parallel \boldsymbol{\rm{H}}_{\rm{ris}}^{H}{(\boldsymbol{\rm{H}}_{\rm{ris}} \boldsymbol{\rm{H}}_{\rm{ris}}^{H} + \kappa \boldsymbol{\rm I}_{K})}^{-1} \parallel},\ K\leq M,\vspace{1.5ex}\\
\dfrac{{(\boldsymbol{\rm{H}}_{\rm{ris}}^{H} \boldsymbol{\rm{H}}_{\rm{ris}} + \kappa \boldsymbol{\rm I}_{M})}^{-1} \boldsymbol{\rm{H}}_{\rm{ris}}^{H}}{\parallel {(\boldsymbol{\rm{H}}_{\rm{ris}}^{H} \boldsymbol{\rm{H}}_{\rm{ris}} + \kappa \boldsymbol{\rm I}_{M})}^{-1} \boldsymbol{\rm{H}}_{\rm{ris}}^{H} \parallel},\ K> M,\\
\end{array} \right.
\label{equi:omega}
\end{equation}
where $\kappa = K {\sigma}_{k}^{2}$ is the regularization parameter assuming the noise power is the same at all \textit{K} users~\cite{mao2022rate}, $\boldsymbol{\rm I}_{K} \in \mathbb{R}^{K\times K}$ and $\boldsymbol{\rm I}_{M} \in \mathbb{R}^{M\times M}$ are the identity matrices.
\vspace{-10pt}

\subsection{Power Consumption Model}
The total power consumption model adopted in this paper is adapted by incorporating those proposed in~\cite{huang2019reconfigurable,tasci2022new,long2021active}, which is given by
\begin{equation}
P_{\rm{total}} = P_{\rm{FBS}} + P_{\rm{RIS}} + KP_{\rm{UE}} + \xi \sum_{k=1}^{K} P_{k},
\label{equi:power consumption}
\end{equation}
where $P_{k}$ is the transmit power for the $k$th UE subject to constraint $\sum_{k=1}^{K} P_{k} \leq P_{t}^{\max}$, and $\xi \triangleq {\nu}^{-1}$ with $\nu$ being the efficiency of the transmit power amplifier; $P_{\rm{FBS}}$, $P_{\rm{RIS}}$, and $P_{\rm{UE}}$ are the power dissipated at FBS, RIS, and each UE, respectively.

From a hardware perspective, the RIS power model characterized by $P_{\rm{RIS}}$ can be further divided into near-constant power consumption of the RIS controller (FPGAs or microcontrollers), power consumption of drive circuits related to RIS complexity and element number, and the power dissipated by reconfigurable elements based on their number and reconfiguration methods. Power consumption features vary significantly across different reconfiguration methods, even with the same resolution and element number.
Therefore, we develop the power consumption model of RIS as
\begin{equation}
\begin{split}
P_{\rm{RIS}} &= P_{\rm{controller}} + P_{\rm{driver}} + P_{\rm{elem}},\\
&= P_{\rm{controller}} + N\cdot P_{\rm{dc}} + P_{\rm{elem}},
\end{split}
\label{equi:P_RIS}
\end{equation}
where $P_{\rm{controller}}$ and $P_{\rm{driver}}$ are the power consumption of the RIS controller and the drive circuits, respectively, and $P_{\rm{elem}}$ is the power consumed by the reconfigurable elements to realize the amplitude and phase adaptation. For better quantitative analysis, we simplify $P_{\rm{driver}}$ to be related to individual reconfigurable elements, assuming that each element has its independent driving circuit for independent controlling. Therefore, its value is the product of the driving circuit power consumption of each element $P_{\rm{dc}}$ and the number of reconfigurable elements $N$.

In~\cite{wang2024reconfigurable}, a measurement-based power consumption model for reconfigurable elements is provided, encompassing three typical reconfiguration methods of RIS. However, its power model for RF switch-based RIS is primarily correlated with the number of elements, making $P_{\rm{elem}}$ in~(\ref{equi:P_RIS}) highly dependant on the specific RF switches that have different numbers of buses and flip-flops. 
For instance, reconfigurable elements based on single multi-bit RF switch and based on multiple 1-bit RF switches have quite different characteristics of power consumption~\cite{rossanese2022designing,rao2022novel}.
To facilitate a more quantitative analysis, we treat each multi-bit RF switch as an assembly of multiple 1-bit RF switches, and construct the power consumption model of unit independently controlled RIS $P_{\rm{elem}}$ under different reconfiguration methods as
\begin{equation}
P_{\rm{elem}}=\left\{
\begin{array}{cl}
P_{\rm{PIN}}\cdot \sum_{i=1}^{N}b_{i}\,,&\rm{PIN\textendash based},\vspace{1.5ex}\\
0\,,&\rm{varactor\textendash based},\vspace{1.5ex}\\
N\cdot R_{b}\cdot P_{\rm{switch}}\,,&\rm{RF~switch\textendash based}.\\
\end{array} \right.
\label{equi:P_element}
\end{equation}
where $b_{i}$ represents the number of the PIN diodes whose states are encoded as `1' in the $i$th reconfigurable element, and $P_{\rm{PIN}}$ is the power consumed to support each PIN diode encoded as `1'; $R_{b}$ is the RIS resolution in bits, and $P_{\rm{switch}}$ is the power consumption per RF switch. For a 2-bit PIN-based reconfigurable element that has two PIN diodes with four states `00', `01', `10', and `11', states `01' and `10' consume the power of $P_{\rm{PIN}}$ while state `11' consume the power of $2P_{\rm{PIN}}$.
For RISs adopting RF switches, the power consumption of switches remains constant regardless of their on-off states, and the number of RF switches also increases the reconfiguration resolution, which will inevitably consume more power. The main difference is that PIN-based RIS has variable element power consumption depending on working states of reconfigurable elements, while the power consumption of RF switch-based RIS is irrespective of their working states.

\section{Problem Formulations and Solutions}
\label{sec:problem formulations and solutions}
In this section, we formulate the near-field EE optimization of the developed RIS-assisted communication system under various reconfiguration methods. By integrating RZF transmit precoding, non-convex optimization-based power allocation, and integer-based discrete optimization of RIS phase configurations, we propose a novel optimization framework that alternatively updates the RIS configurations and BS power allocation to achieve maximal EE in RIS-assisted near-field communication systems. In the following parts, we present our alternating optimization framework with detailed optimization approaches for RIS configuration and power allocation successively and give the analysis of computational complexity.

\subsection{Problem Formulations}
The received signal-to-interference-plus-noise ratio (SINR) $\gamma_{k}$ of $k$th UE in this system can be formulated as
\begin{equation}
\gamma_{k}({{\boldsymbol {\Phi}}\,, \boldsymbol {\rm p}}) = \dfrac{P_k{\mid \boldsymbol{\rm{h}}_{\rm{ris}\it{,k}}~\boldsymbol{\rm{\omega}}_{k} \mid}^{2}}{\sum^{K}_{j=1,j\neq k}P_j{\mid \boldsymbol{\rm{h}}_{\rm{ris}\it{,k}}~\boldsymbol{\rm{\omega}}_{j} \mid}^{2}+\sigma^{2}_{k}}.
\label{equi:SINR_k}
\end{equation}
where $\boldsymbol {\rm p} \triangleq [P_1, P_2, ..., P_K]$.
Then, based on (\ref{equi:SINR_k}), the SE in bps/Hz of this system is given by
\begin{equation}
\eta_{\rm{SE}}({{\boldsymbol {\Phi}}\,, \boldsymbol {\rm p}}) = \sum_{k=1}^{K} {\rm log}_{2}(1+\gamma_{k}({{\boldsymbol {\Phi}}\,, \boldsymbol {\rm p}})).
\label{equi:sum rate}
\end{equation}
The EE in bit-per-Joule is defined as
\begin{equation}
\eta_{\rm{EE}} \triangleq \dfrac{\rm{BW} \cdot \eta_{\rm{SE}}}{P_{\rm{total}}},
\label{equi:EE_1}
\end{equation}
where $\rm{BW}$ is the bandwidth of transmission. By combining (\ref{equi:power consumption}), (\ref{equi:P_RIS}), (\ref{equi:SINR_k}), and (\ref{equi:sum rate}), (\ref{equi:EE_1}) given above can be further expressed as (\ref{equi:EE_2}) at the top of the next page.

\begin{figure*}[ht!]
\begin{equation}
\eta_{\rm{EE}}({{\boldsymbol {\Phi}}\,, \boldsymbol {\rm p}}) = \frac{{\rm BW} \cdot \sum_{k=1}^{K} {\rm log}_{2}(1+\frac{P_k{\mid \boldsymbol{\rm{h}}_{\rm{ris}\it{,k}}~\boldsymbol{\rm{\omega}}_{k} \mid}^{2}}{\sum^{K}_{j=1,j\neq k} P_j {\mid \boldsymbol{\rm{h}}_{\rm{ris}\it{,k}}~\boldsymbol{\rm{\omega}}_{j} \mid}^{2}+\sigma^{2}_{k}})}{P_{\rm{FBS}} + P_{\rm{controller}} + N\cdot P_{\rm{dc}} + P_{\rm{elem}} + KP_{\rm{UE}} + \xi \sum_{k=1}^{K} P_{k}}
\label{equi:EE_2}
\end{equation}
\hrulefill
\vspace{-10pt}
\end{figure*}

Based on the continuous FBS transmit precoding and power allocation, and the discrete reconfiguration of RIS phase arrangement, the optimal EE of RIS-assisted near-field communications can be formulated as
\begin{IEEEeqnarray}{ccl}\label{P_1}
{\rm\mathcal P1:} \,\, &\underset{{\boldsymbol {\Phi}}\,, \boldsymbol {\rm p}}{\max}\ & \eta_{\rm{EE}}({{\boldsymbol {\Phi}}\,, \boldsymbol {\rm p}})
\\
	&\text{s.t.} &\theta^{\mathrm{disc}}_{n} \in S \,, \forall n \in \{1,2,...,N\}, \IEEEyessubnumber \label{P_1a}
    \\
    && \sum_{k=1}^K P_{k} \le P_t^{\max},\IEEEyessubnumber \label{P_1b}
	\\
	&&P_{k}\ge 0\,,  \forall k \in \{1,2,...,K\}, \IEEEyessubnumber \label{P_1c}
\end{IEEEeqnarray}
where the discrete phase $\theta^{\mathrm{disc}}_{n}$ of $n$th reconfigurable element has a value space $S = \left\{\frac{2\pi}{2^{R_b}}(i+\frac{1}{2}) \mid i=0,1,..., 2^{R_b}-1\right\}$, determined by the RIS resolution $R_b$, and the total transmit power $\sum_{k=1}^K P_{k}$ for all $K$ UEs must not exceed the maximum allowable transmit power $P_t^{\max}$.
\vspace{-10pt}

\subsection{Integer-PSO Algorithm for Reconfiguration}
\begin{figure}[t]
    \centering
    \includegraphics[width=0.8\columnwidth]{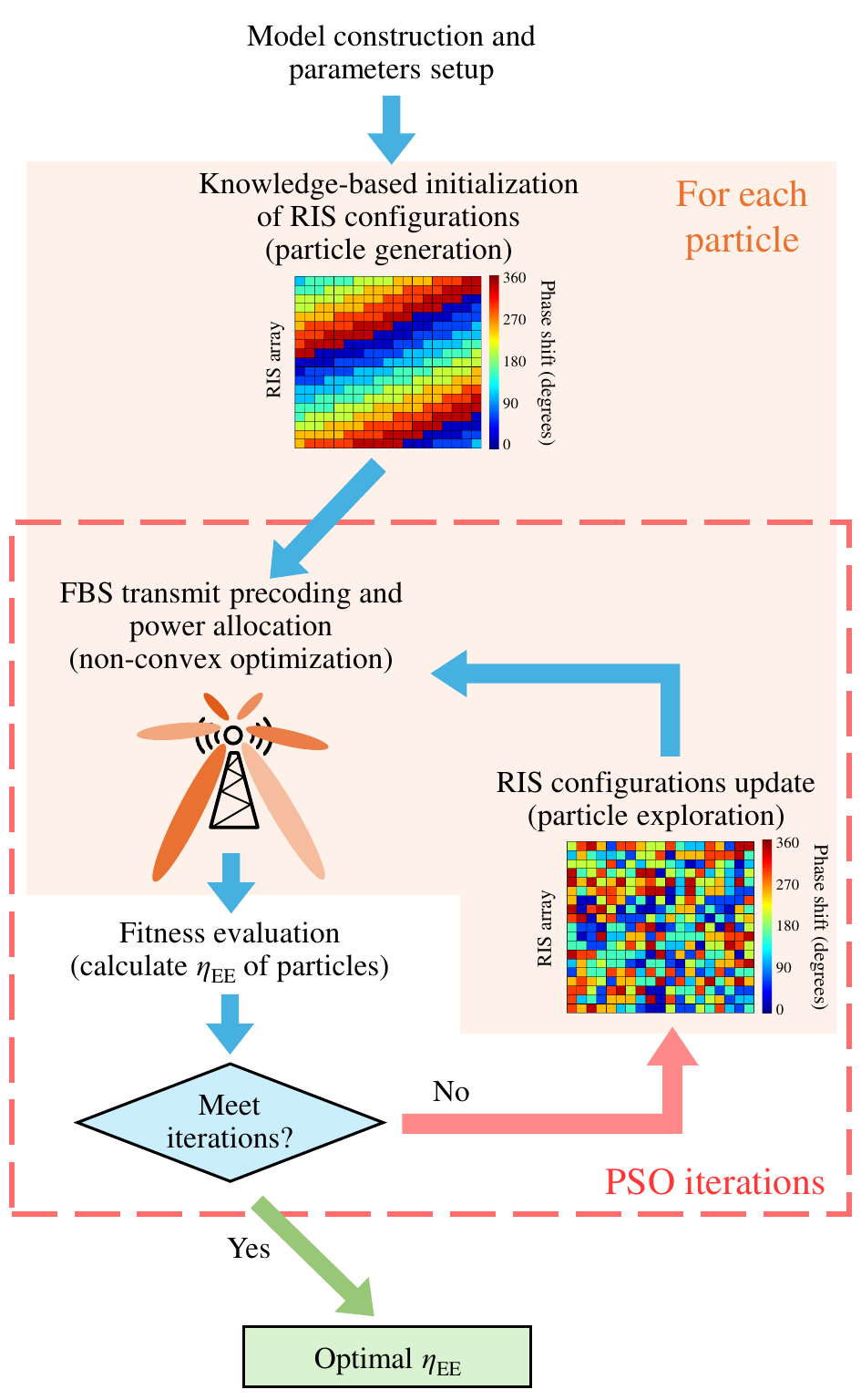}
    \vspace{-5pt}
    \caption{Alternative optimization framework comprised of the proposed integer-PSO-based RIS phase shift optimization and Dinkelbach algorithm-based FBS transmit power allocation.}
    \label{fig:PSO_Convex}
\vspace{-10pt}
\end{figure}

In this subsection, we employ integer PSO for RIS phase shift reconfigurations. PSO~\cite{488968} is a typical heuristic algorithm designed to efficiently explore high-dimensional spaces and search for the global optimum through distributed particles operating under predefined policies. 
Although the near-field conditions themselves may not directly affect the formulated EE optimization, certain characteristics associated with near-field communications, such as the adoption of extremely large RIS arrays and the resulting significantly higher proportion of RIS power consumption in the system’s total power consumption, increase the complexity of near-field EE optimization. Additionally, the discrete RIS phase shift optimization is a high-dimensional mixed-integer nonlinear programming (MINLP) problem. The practical power consumption model of PIN-based RISs further introduces a coupling between the RIS power consumption and real-time RIS configurations, which complicates the problem by making both the numerator and denominator of the formulated problem dependent on the RIS configurations. In these circumstances, PSO offers greater flexibility than convex optimization by avoiding case-by-case transformation design, relaxations, or quantization~\cite{10361836} and higher optimization efficiency than learning-based methods by avoiding the need for extensive training data acquisition and time-greedy training~\cite{wang2024drl}.

In~\cite{wang2024integer}, integer PSO has been demonstrated to achieve optimal radiation pattern synthesis for RIS-assisted communications with high computing efficiency and low time complexity by optimizing the arrangement of limited and discrete phase shifts on the RIS array.
In this work, we further improve this integer phase shift optimization algorithm and integrate it with RZF precoding and non-convex power allocation to address the EE maximization problem of RIS-assisted near-field communications systems with cascaded channels. As shown in Fig.~\ref{fig:PSO_Convex}, the precoding and power allocation are applied to each particle and inserted into each integer-PSO step, ensuring that the EE of each particle in each step is optimal under its current RIS phase configuration. Considering that RIS reconfigurability can be achieved by various methods, each having specific power consumption features, we also analyze the performance variations of the RIS-assisted near-field communication systems when different reconfiguration methods and bit resolutions are employed.

In our proposed solution, the position of each particle corresponds to a specific phase shift arrangement of $\sqrt{N}\times \sqrt{N}$ rectangular RIS that each reconfigurable element is represented by an integer value of the phase shift state depending on reconfiguration resolution $R_b$. Unlike the random initialization of particles, we generate the initial particles by utilizing the average of random weights of the calculated optimal RIS phase arrangements for individual UEs. The initialized continuous phase arrangement of the $p$th particle is given by
\begin{equation}
\boldsymbol{\theta}_p^{\mathrm{cont}} \triangleq \dfrac{1}{M}\sum_{m=1}^{M} r^{\rm{in}}_{p,m}\cdot{\boldsymbol{\varphi}^{\rm{in}}_{N,m}} + \dfrac{1}{K}\sum_{k=1}^{K} r^{\rm{re}}_{p,k}\cdot{\boldsymbol{\varphi}^{\rm{re}}_{k,N}},
\label{equi:Initial_RIS_continuous}
\end{equation}
where $r^{\rm{in}}_{p,m}$ and $r^{\rm{re}}_{p,k}$ are two Gaussian distributed random variables with a mean of 1 as the random weights for the superposition; ${\boldsymbol{\varphi}^{\rm{in}}_{N,m}}=[\varphi^{\rm{in}}_{1,m}, \varphi^{\rm{in}}_{2,m},..., \varphi^{\rm{in}}_{N,m}]$ and ${\boldsymbol{\varphi}^{\rm{re}}_{k,N}}=[\varphi^{\rm{re}}_{k,1}, \varphi^{\rm{re}}_{k,2},..., \varphi^{\rm{re}}_{k,N}]$ are the incident and reflected phase responses of the cascaded channel from $m$th FBS antenna to $k$th UE via RIS, respectively.
After that, discrete phase $\boldsymbol{\theta}_p^{\mathrm{disc}}\in \mathbb{C}^{1\times N}$ rounded from $\boldsymbol{\theta}_p^{\mathrm{cont}}$ is further mapped to an integer vector with all the entries between $[0,2^{R_b}-1]$, then, converted to be the rectangular initialized position of the $p$th particle as ${\rm {\bf x}}_p^{[0]}\in \mathbb{C}^{\sqrt{N}\times \sqrt{N}}$, for the RIS with $R_b$ resolution.

By introducing the knowledge-based generation of initial particles and the discard rates for faster and better convergence, velocity ${\rm {\bf v}}_p\in \mathbb{C}^{\sqrt{N}\times \sqrt{N}}$ and position ${\rm {\bf x}}_p\in \mathbb{C}^{\sqrt{N}\times \sqrt{N}}$ of $p$th particle at the $t$th step are updated by
\begin{equation}
\begin{split}
    {\rm {\bf v}}_p^{[t]}\triangleq~&w\cdot {\rm {\bf v}}_p^{[t-1]}+c_1\cdot r_{1,p}\cdot {\rm {\bf d}}_1\cdot ({\rm {\bf Pm}}_p^{[t-1]}-{\rm {\bf x}}_p^{[t-1]})\\
    &+c_2\cdot r_{2,p}\cdot {\rm {\bf d}}_2\cdot ({\rm {\bf Gm}}^{[t-1]}-{\rm {\bf x}}_p^{[t-1]}),
\end{split}
\label{equi:velocity}
\end{equation}
and
\begin{equation}
{\rm {\bf x}}_p^{[t]}\triangleq{\rm round}({\rm {\bf x}}_p^{[t-1]}+{\rm {\bf v}}_p^{[t-1]}),
\label{equi:position}
\end{equation}
where $w$ is the inertia factor; $c_1$ and $c_2$ are the cognitive acceleration coefficient and social acceleration coefficient, respectively; $d_1$ and $d_2$ are the cognitive discard rate and social discard rate, respectively. We propose the discard rates to decelerate convergence because limited phase states, especially in low reconfiguration resolutions, make the particles easily get stuck in local optimums.
Besides, $r_{1,p} \in (0,1)$ and $r_{2,p} \in (0,1)$ are uniform random variables to introduce the randomness for the search; ${\rm {\bf Pm}}_p\in \mathbb{C}^{\sqrt{N}\times \sqrt{N}}$ is the cognitive best position of particle~\textit{p} while ${\rm {\bf Gm}}\in \mathbb{C}^{\sqrt{N}\times \sqrt{N}}$ is the social best position in history.

To ensure valid convergence, we impose the restriction of $[-1,1]$ for 1-bit resolution and $[-2^{R_b}/4,2^{R_b}/4]$ for resolutions greater than 2 bits on updating ${\rm {\bf v}}_p$ to control the converging rate. Meanwhile, initial velocity ${\rm {\bf v}}_p^{[0]}$ is generated randomly within the velocity restriction. Then, the algorithm rounds the sum of ${\rm {\bf v}}_p$ and ${\rm {\bf x}}_p$ to the nearest integer so that the optimization and adjustment of the reconfigurable elements' phase shifts are always in the integer parameter space. For the RISs with $R_b$-bit resolution, the solution space of each entry of ${\rm {\bf x}}_p$ is the integers between $[0,2^{R_b}-1]$. Therefore, for the entries in updated ${\rm {\bf x}}_p$ that exceed the solution space, their values will be normalized, ensuring that the phase shifts of all reconfigurable elements always remain within the range of $[0,360^{\circ})$.

After updating the RIS configuration through the guided search by particles, the power consumption $P_{\rm controller}$, $P_{\rm dc}$, and $P_{\rm elem}$, that related to current RIS configuration, and cascaded channel matrix $\boldsymbol{\rm{H}}_{\rm{ris}}$ can be determined. Meanwhile, the precoding vectors $\boldsymbol{\rm{\omega}}_{k}\,,\forall k \in \{1,...,K\}$, can also be calculated by RZF precoding as shown in~(\ref{equi:omega}).
Then, we adopt a two-layer non-convex iterative algorithm for each particle to optimize the FBS power allocation under their corresponding current RIS configurations, aiming to further increase the system's EE.
\vspace{-10pt}

\subsection{Dinkelbach-IQT Algorithm for Power Allocation}
In this subsection, we focus on the power allocation scheme after obtaining the optimal RIS phase configuration for each particle in each integer-PSO step. Specifically, we first rewrite the SINR expression given in~(\ref{equi:SINR_k}) as
\begin{equation}\label{PA1}
\gamma _k({\boldsymbol {\rm p}})=\frac{P_k\zeta _{k,k}}{\sum_{j=1,j\ne k}^K{P_j\zeta _{k,j}}+\sigma _{k}^{2}},
\end{equation}
where $\zeta_{k,k} = \mid \mathbf{h}_{\mathrm{ris}, k}~\boldsymbol{\omega}_k \mid^2$ and $\zeta_{k,j} = \mid \mathbf{h}_{\mathrm{ris}, k}~\boldsymbol{\omega}_j \mid^2$. 
Since $P_{\rm{FBS}}$, $P_{\rm{controller}}$, $N\cdot P_{\rm{dc}}$, $P_{\rm{elem}}$, and $KP_{\rm{UE}}$ are constant for the given cascaded system and RIS configuration, we define $P_{\rm{fixed}} \triangleq P_{\rm{FBS}} + P_{\rm{controller}} + N\cdot P_{\rm{dc}} + P_{\rm{elem}} + KP_{\rm{UE}}$, and $\eta_{\rm{EE}}$ in (\ref{equi:EE_2}) can be rewritten as
\begin{equation}\label{PA_EE}
\eta_{\rm{EE}}({\boldsymbol {\rm p}})=\frac{{{\rm BW}\cdot \sum_{k=1}^K{\log _2\left( 1+\gamma _k({\boldsymbol {\rm p}}) \right)}}}{P_{\mathrm{fixed}}+\xi \sum_{k=1}^K{P_k}},
\end{equation}
Then, the power allocation problem can be formulated as
\begin{IEEEeqnarray}{ccl}\label{PA2}
	\ {\rm\mathcal P2:} \,\, &\underset{\boldsymbol {\rm p}}{\max}\quad & \eta_{\rm{EE}}({\boldsymbol {\rm p}})
	\\
	&\text{s.t.} & \sum_{k=1}^K P_{k} \le P_t^{\max}, \IEEEyessubnumber \label{PA2a}
	\\
	&&P_{k}\ge 0,  \forall k \in \{1,2,...,K\}.\IEEEyessubnumber \label{PA2b}
\end{IEEEeqnarray}
Problem $\mathcal{P}2$ is a non-convex fractional programming problem and, thus, the Dinkelbach algorithm-based optimization framework can be effectively employed to obtain a local optimum with a low complexity.

To begin with, according to the principle of the Dinkelbach algorithm~\cite{dinkelbach1967nonlinear}, maximal EE $\eta_{\rm{EE}}^*$ satisfies
\begin{align}\label{PA3}
& \underset{{\boldsymbol {\rm p}}}{\max} \left\{{\rm BW}\cdot \sum_{k=1}^K{\log _2\left( 1+\gamma _k({\boldsymbol {\rm p}}) \right)} - \eta_{\rm{EE}}^*(P_{\mathrm{fixed}}+\xi \sum_{k=1}^K{P_k})\right\} \notag 
\\
& = {\rm BW}\cdot \sum_{k=1}^K{\log _2\left( 1+\gamma _k({\boldsymbol {\rm p}}^*) \right)} - \eta_{\rm{EE}}^*(P_{\mathrm{fixed}}+\xi \sum_{k=1}^K{P_k^*})  \notag 
\\
& = 0,
\end{align}
where ${\boldsymbol {\rm p}}^*$ is the optimal power allocation coefficient and can be obtained by solving the problem formulated in (\ref{PA3}).

However, $\eta_{\rm{EE}}^*$ as the result of the objective function of problem $\mathcal{P}2$, is still unknown by solving (\ref{PA3}). In this regard, we employ the iterative Dinkelbach algorithm to determine $\eta_{\rm{EE}}^*$. Specifically, at the $n$th iteration, we set $\eta_{\rm{EE}}^{[n]}$ as follows
\begin{equation}\label{PA4}
\eta _{\mathrm{EE}}^{[n]}=\frac{\small{{\rm BW}\cdot \sum_{k=1}^K{\log _2\left( 1+\gamma _k\left( {\boldsymbol {\rm p}}^{[n-1]} \right) \right)}}}{P_{\mathrm{fixed}}+\xi \sum_{k=1}^K{P_{k}^{[n-1]}}},
\end{equation}
where ${\boldsymbol {\rm p}}^{[n-1]}$ denotes the result of the $(n-1)$-th iteration by the Dinkelbach algorithm, and ${\boldsymbol {\rm p}}^{[0]}$ refers to a given initial value. Then, solving $\mathcal{P}2$ is equivalent to solving the following problem
\begin{IEEEeqnarray}{ccl}\label{PA5}
\ {\rm\mathcal P3:} \,\, & \underset{{\boldsymbol {\rm p}}}{\max}\quad & J({\boldsymbol {\rm p}})
\\
&\text{s.t.} & \text{(\ref{PA2a})}, \text{(\ref{PA2b})}, \IEEEnonumber
\vspace{-5pt}
\end{IEEEeqnarray}
where $J({\boldsymbol {\rm p}})$ is given by
\begin{equation}\label{PAQ1}
J({\boldsymbol {\rm p}}) = {\rm BW} \cdot \sum_{k=1}^K{\log _2\left( 1+\gamma _k({\boldsymbol {\rm p}}) \right)} - \eta_{\rm{EE}}^{[n]}(P_{\mathrm{fixed}}+\xi \sum_{k=1}^K{P_k}).
\end{equation}

However, problem $\mathcal{P}3$ is a non-convex problem due to the linear fractional expression of $\gamma_k({\boldsymbol {\rm p}})$. Therefore, an iterative quadratic transform (IQT) method is utilized to tackle its non-convexity. We first formulate the dual problem of $\mathcal{P}3$ as follows \cite{shen2018fractional1}
\begin{IEEEeqnarray}{ccl}\label{PA6}
\ {\rm\mathcal P4:} \,\, & \underset{{\boldsymbol {\rm p}}, \mathbf{\Gamma}}{\max}\quad & \vartheta \sum_{k=1}^K{\ln \left( 1+\Gamma _k \right)}-\vartheta \sum_{k=1}^K{\Gamma _k}+F\left( \mathbf{{\boldsymbol {\rm p}} , \Gamma}\right)
\\
&\text{s.t.} & \text{(\ref{PA2a})}, \text{(\ref{PA2b})}, \IEEEnonumber
\end{IEEEeqnarray}
where $\mathbf{\Gamma} = [\Gamma_1,... ,\Gamma_K] \in \mathbb{R}^{K \times 1}$ denotes a sequence of non-negatively auxiliary variables, and $\vartheta = 1/\ln (2)$; $F\left( \mathbf{{\boldsymbol {\rm p}} , \Gamma}\right)$ is given by
\begin{equation}\label{PA7}
F\left( {\boldsymbol {\rm p}},\mathbf{\Gamma } \right) \!=\! \sum_{k=1}^K{\frac{\vartheta \left( 1+\Gamma _k \right) P_k\zeta _{k,k}}{\sum_{j=1}^K{P_j\zeta _{k,j}}+\sigma _{k}^{2}}}-\eta _{\mathrm{EE}}^{[n]}\left(\! P_{\mathrm{fixed}}\!+\!\xi \sum_{k=1}^K{P_k} \!\right).
\end{equation}

Then, we update $\mathbf{\Gamma}$ and ${\boldsymbol {\rm p}}$ separately. For a given initial ${\boldsymbol {\rm p}}^{[t]}$, problem $\mathcal{P}4$ is a non-constrained convex problem with respect to $\mathbf{\Gamma}$, and the optimal solution can be achieved by
\begin{equation}\label{PA8}
\Gamma _{k}^{[t+1]}=\frac{P_{k}^{[t]}\zeta _{k,k}}{\sum_{j=1,j\ne k}^K{P_{j}^{[t]}\zeta _{kj}}+\sigma _{k}^{2}}.
\end{equation}
Based on $\mathbf{\Gamma}^{[t+1]}$, we then update ${\boldsymbol {\rm p}}$ by solving the following problem
\begin{IEEEeqnarray}{ccl}\label{PA9}
\ {\rm\mathcal P5:} \,\, & \underset{{\boldsymbol {\rm p}}}{\max}\quad & F({\boldsymbol {\rm p}}, \mathbf{\Gamma}^{[t+1]})
\\
&\text{s.t.} & \text{(\ref{PA2a})}, \text{(\ref{PA2b})}. \IEEEnonumber
\end{IEEEeqnarray}
By applying quadratic transform \cite{8314727}, problem $\mathcal{P}5$ can be reformulated as
\begin{IEEEeqnarray}{ccl}\label{PA10}
\ {\rm\mathcal P6:} \,\, & \underset{{\boldsymbol {\rm p}}, \mathbf{y}}{\max}\quad & G({\boldsymbol {\rm p}}, \mathbf{\Gamma}^{[n+1]}, \mathbf{y})
\\
&\text{s.t.} & \text{(\ref{PA2a})}, \text{(\ref{PA2b})}, \IEEEnonumber
\vspace{-5pt}
\end{IEEEeqnarray}
where $\mathbf{y} = [y_1, ..., y_K] \in \mathbb{R}^{K\times 1}$ is a sequence of non-negative auxiliary variables, and $G({\boldsymbol {\rm p}}, \mathbf{\Gamma}^{[t+1]}, \mathbf{y})$ is given by
\begin{align}\label{PA11}
G\left( {\boldsymbol {\rm p}},\Gamma ^{[t+1]},\mathbf{y} \right) = &\sum_{k=1}^K{2y_k\sqrt{\vartheta \left( 1+\Gamma _{k}^{[t+1]} \right) P_k\zeta _{k,k}}}
\notag \\
&-\sum_{k=1}^K{y_{k}^{2}\left( \sum_{j=1}^K{P_j\zeta _{k,j}+\sigma _{k}^{2}} \right)}
\notag \\
&-\eta _{\mathrm{EE}}^{[n]}\left( P_{\mathrm{fixed}}+\xi \sum_{k=1}^K{P_k} \right).
\end{align}

Similarly, we update ${\boldsymbol {\rm p}}$ and $\mathbf{y}$ separately to tackle problem $\mathcal{P}6$. For a given ${\boldsymbol {\rm p}}^{[t]}$, problem $\mathcal{P}6$ is convex with respect to the collection of variables $\mathbf{y}$. Thus, solving $\partial G\left( {\boldsymbol {\rm p}},\Gamma ^{[t+1]},\mathbf{y} \right) /\partial y_k=0$ yields
\begin{equation}\label{PA12}
y_{k}^{[t+1]}=\frac{\sqrt{\vartheta ( 1+\Gamma _{k}^{[t+1]} ) P_{k}^{[t]}\zeta _{k,k}}}{\sum_{j=1}^K{P_{j}^{[t]}\zeta _{k,j}}+\sigma _{k}^{2}}.
\vspace{-5pt}
\end{equation}
With $\mathbf{\Gamma}^{[t+1]}$ and $\mathbf{y}^{[t+1]}$, problem $\mathcal{P}6$ is convex over ${\boldsymbol {\rm p}}$. The following lemma describes the optimality of power allocation coefficient ${\boldsymbol {\rm p}}$:
\par \textit{Lemma 1:} The optimal power allocation coefficient ${\boldsymbol {\rm p}}^{[t+1]}$ satisfies\footnote{Note that the problem $\mathcal{P}6$ w.r.t. $\mathbf{p}$ can be solved directly using CVX. However, we propose Lemma 1 to reduce the solving complexity.}
\begin{equation}\label{PA13}
P_{j}^{[t+1]}(\rho)=\frac{\left( y_{k}^{[t+1]} \right) ^2\vartheta \left( 1+\Gamma _{k}^{[t+1]} \right) \zeta _{k,k}}{\left( \rho ^*+\eta _{\mathrm{EE}}^{[n]}\xi +\sum_{j=1}^K{\left( y_{j}^{[t+1]} \right) ^2\zeta _{j,k}} \right) ^2},
\end{equation}
where $\rho$ is given by
\begin{equation}\label{PA14}
\begin{cases}
	\rho = 0, \, \text{if} \sum_{k=1}^K{P_{k}^{[t+1]}\left( \rho \right)}\le P_{t}^{\max},\\
	\rho = \rho^*, \, \text{otherwise},
\end{cases}
\end{equation}
in which $\rho^*$ is the solution to $P_t^{\max} - \sum_{k=1}^KP_k^{[t+1]}(\rho)=0$, which can be obtained via the bisection method.
\par \textit{Proof:} See Appendix. 

The convergence proof of the IQT algorithm described above is provided in \cite{shen2018fractional1}. Therefore, under a given tolerance $\epsilon_{\rm{inner}}$, by iteratively updating $\mathbf{\Gamma}$, $\mathbf{y}$, and ${\boldsymbol {\rm p}}$ based on (\ref{PA8}), (\ref{PA12}), and (\ref{PA13}) until $\lvert J({\boldsymbol {\rm p}}^{[t+1]})-J({\boldsymbol {\rm p}}^{[t]})\rvert \le \epsilon_{\rm{inner}}$ is satisfied, a local optimal solution to problem $\mathcal{P}3$ can be obtained, with $P_k^{[n+1]} = P_k^{[t+1]}$. Similarly, the convergence of the Dinkelbach transform is established in \cite{dinkelbach1967nonlinear}. Then, $P_k^{[n+1]}$ can be used to proceed with the outer-layer Dinkelbach algorithm until local optimal $\eta_{\rm{EE}}^*$ is achieved when $\lvert J({\boldsymbol {\rm p}}^{[n+1]}) \rvert \le \epsilon_{\rm{outer}}$. The detailed process of the iterative Dinkelbach algorithm and the IQT method are summarized in Algorithm \ref{Algorithm 1}.

\subsection{Complexity Analysis}
\vspace{-5pt}
As presented above, we integrate the power allocation optimization in each iteration of RIS phase arrangement optimization. Therefore, the computational complexity should be the multiplication of integer-PSO and Dinkelbach-IQT algorithm. Assuming that there are $N_{\rm p}$ particles and $I_{\rm PSO}$ iterations in our integer-PSO setup, the computational complexity of the proposed integer-PSO scheme is thus $\mathcal{O}(N_{\rm p} I_{\rm PSO})$.
Nonlinear integer optimization problems are NP-hard\cite{9497709}, and the possible configurations of RISs are $2^{R_b N}$, making exhaustive search infeasible. Therefore, our proposed integer-PSO algorithm provides an effective approach to optimize RIS-assisted communications with a high number of reconfigurable elements $N$ and phase states $2^{R_b}$ while maintaining limited computational complexity.
Next, we analyze the complexity of the power control scheme. The complexity of Dinkelbach algorithm is $\mathcal{O}(\epsilon_{outer}^{-2}\ln2)$\cite{dinkelbach1967nonlinear}. In each outer-layer iteration, where the IQT algorithm is performed, if we use the bisection method to determine $\rho^*$, the complexity of IQT algorithm is $\mathcal{O}(I_{\rm IQT}K\ln(1/\epsilon_{inner}))$, where $I_{\rm IQT}$ denotes the number of iterations of IQT method. Therefore, the overall complexity of the power allocation scheme is $\mathcal{O}(I_{\rm IQT}K)$ and the complexity of the whole alternating optimization is $\mathcal{O}(N_{\rm p} I_{\rm PSO} I_{\rm IQT}K)$.

\begin{algorithm}[t]
\caption{Power allocation scheme}\label{Algorithm 1}
\LinesNumbered
Initialize outer-layer iteration tolerance $\epsilon_{\rm{outer}}$ and index $n=0$. Initialize ${\boldsymbol {\rm p}}^{[n]}$.\\
\Repeat{$\mid J({\boldsymbol {\rm p}}^{[n]}) \mid \le \epsilon_{\rm{outer}}$}
    {
    Update $\eta_{\rm{EE}}^{[n]}$ based on (\ref{PA4}). \\
    Initialize inner-layer iteration tolerance $\epsilon_{\rm{inner}}$ and index $t=0$. 
    Initialize ${\boldsymbol {\rm p}}^{[t]} = {\boldsymbol {\rm p}}^{[n]}$ and $J({\boldsymbol {\rm p}}^{[t]})$. \\
    \Repeat{$\mid J({\boldsymbol {\rm p}}^{[t]}) - J({\boldsymbol {\rm p}}^{[t-1]}) \mid \le \epsilon_{\rm{inner}}$}
        {
        Set $t \gets t + 1$. \\
        Update $\mathbf{\Gamma}^{[t]}$ based on (\ref{PA8}). \\
        Update $\mathbf{y}^{[t]}$ based on (\ref{PA12}). \\
        Update ${\boldsymbol {\rm p}}^{[t]}$ based on (\ref{PA13}).\\
        Compute $J({\boldsymbol {\rm p}}^{[t]})$ based on (\ref{PAQ1}). \\
	}
    Set $n \gets n+1$.\\
    Set ${\boldsymbol {\rm p}}^{[n]} = {\boldsymbol {\rm p}}^{[t]}$\\
    }
\textbf{return} $\mathrm{SR}_{\mathrm{MMF}}^{\mathrm{I}}$ and $\mathrm{SR}^{\mathrm{I}}$.
\end{algorithm}
\vspace{-10pt}

\section{Numerical Results and Analysis}
\vspace{-5pt}
\label{sec:numerical results and snalysis}
In this section, we first specify our simulation scenario and setup of the RIS-assisted near-field communication system. Then, we present the numerical results obtained under various conditions, including different RIS reconfiguration methods, different numbers and resolutions of reconfigurable elements, and different maximum transmit power at FBSs. Through a comprehensive evaluation of these factors, we verify that the optimal RIS setup that maximizes the EE of RIS-assisted near-field systems in different situations can be yielded by the proposed alternating optimization framework.

\subsection{Simulation Setup}
\begin{figure}[t]
    \centering
    \includegraphics[width=0.6\columnwidth]{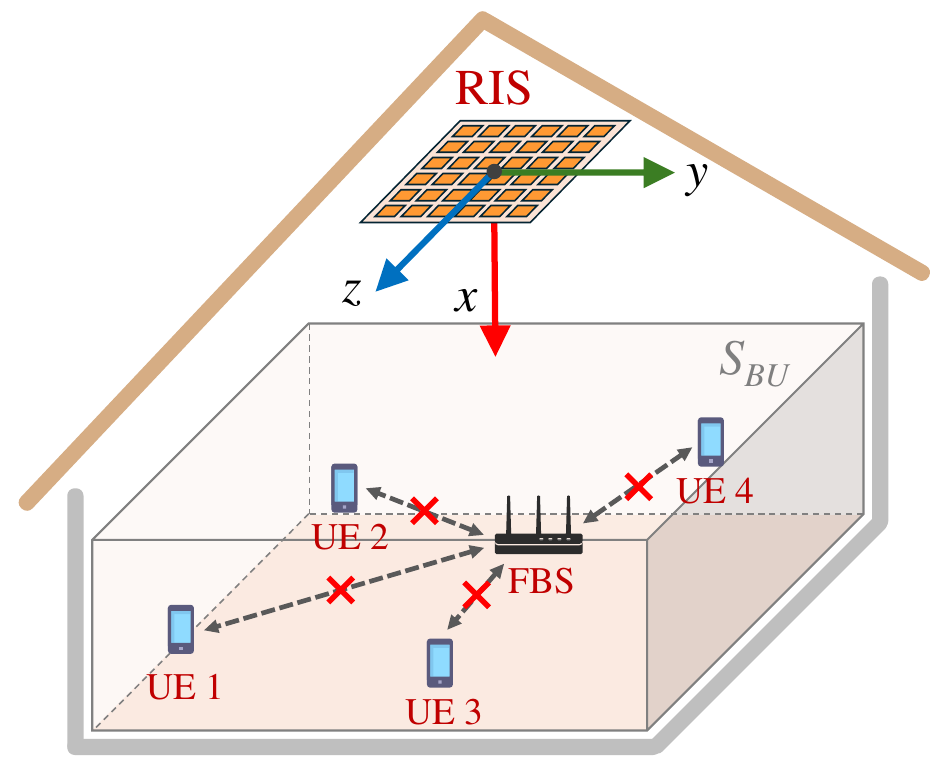}
    \vspace{-5pt}
    \caption{Illustration of the simulated near-field scenario, where an RIS is deployed at the top of the indoor space, while FBS and UEs are randomly distributed within the solving space $S_{BU}$.}
    \label{fig:setup}
\vspace{-10pt}
\end{figure}

\begin{table*}[ht]
\centering
\renewcommand{\arraystretch}{1.3}
\caption{Parameters of simulation setup}
\footnotesize
\label{table:simulation_parameter}
\begin{tabular}{|l|c|l|c|} 
\hline
Parameters & Values & Parameters & Values\\ 
\hline
Number of FBS elements & $M=8$ & Rician factor of RIS-UE channel & $\varepsilon_{h}=5$\\
Number of RIS elements & $N\in[10,20,...,60]^2$ & Rician factor of FBS-RIS channel & $\varepsilon_{G}=5$\\
Number of UEs & $K=4$ & Maximum transmit power of FBS & $P_{t}^{\max}=25$ dBm \cite{9082809}\\
Carrier frequency & $f=5.25$ GHz & Power dissipated at FBS & $P_{\rm FBS}=30$ dBm \cite{9246508}\\
Wavelength & $\lambda=0.0571$ m & Power dissipated at UE & $P_{\rm UE}=10$ dBm \cite{yang2021energy}\\
Element spacing of FBS & $d_{\rm FBS}=\lambda/2$ & Amplifier efficiency & $\nu=0.8$ \cite{yang2021energy}\\
Element spacing of RIS & $d_{\rm RIS}=\lambda/4$ & Power consumption of RIS controller & $P_{\rm controller}=10$ mW \cite{pei2021ris}\\
Resolution of RIS elements & $R_b\in[1,2,...,10]$ bit & Power consumption of drive circuit (varactor) & $P_{\rm{dc}}=4$ mW \cite{pei2021ris}\\
Coordinate of the RIS center & (0, 0, 0) m & Power consumption of drive circuit (PIN \& switch) & $P_{\rm{dc}}=0.01$ mW \cite{wang2024reconfigurable}\\
Distribution region of FBS \& UEs & $\mathcal{R}_x=[4, 6]$ m & Power consumption of PIN encoded as `1' & $P_{\rm PIN}=1.25$ mW \cite{wang2024reconfigurable}\\
Distribution region of FBS \& UEs & $\mathcal{R}_y=[-8, 8]$ m & Power consumption of RF switch & $P_{\rm switch}=0.5$ mW \cite{wang2024reconfigurable}\\
Distribution region of FBS \& UEs & $\mathcal{R}_z=[-8, 8]$ m & Number of particles in integer-PSO & $N_{\rm p} = 100$\\
Transmission bandwidth & BW $=10$ MHz & Number of iteration steps in integer-PSO & $I_{\rm PSO} = 100$\\
\hline
\end{tabular}
\renewcommand{\arraystretch}{1.00}
\vspace{-10pt}
\end{table*}

Fig.~\ref{fig:setup} illustrates the simulation scenario in our study. Here, $\sqrt{N}\times\sqrt{N}$ RIS is assumed to be deployed on the ceiling of residence to establish cascaded channels, aiming to effectively mitigate signal blockage caused by indoor objects. Constructing the three-dimensional Cartesian coordinate system with its origin at the center of the RIS array. The solution space $S_{BU}$, where the \textit{M}-element FBS and four single-antenna UEs are randomly deployed, can be expressed as
\begin{equation}
    S_{BU} \triangleq \{(s_x, s_y, s_z)\mid s_x\in \mathcal{R}_x, s_y\in \mathcal{R}_y, s_z\in \mathcal{R}_z\}.
\end{equation}
As described in Subsection~\ref{subsec:signal model}, the direct channels between the FBS and all UEs are blocked.

The simulation parameters of this RIS-assisted near-field communication scenario are listed in Table~\ref{table:simulation_parameter}. Its power consumption-related parameters are set as per the studies reported in \cite{huang2019reconfigurable,tasci2022new,pei2021ris} and the practical measured values obtained in \cite{wang2024reconfigurable,zhang2019breaking,pei2021ris,rossanese2022designing}. To better reveal the performance when applying different reconfiguration methods and various RIS specifications, the simulation results are based on the average of 200 randomly generated position sets of FBS and UEs.

\subsection{Simulation Results}
\subsubsection{Convergence of proposed algorithm}
\begin{figure}[t]
    \centering
    \includegraphics[width=0.65\columnwidth]{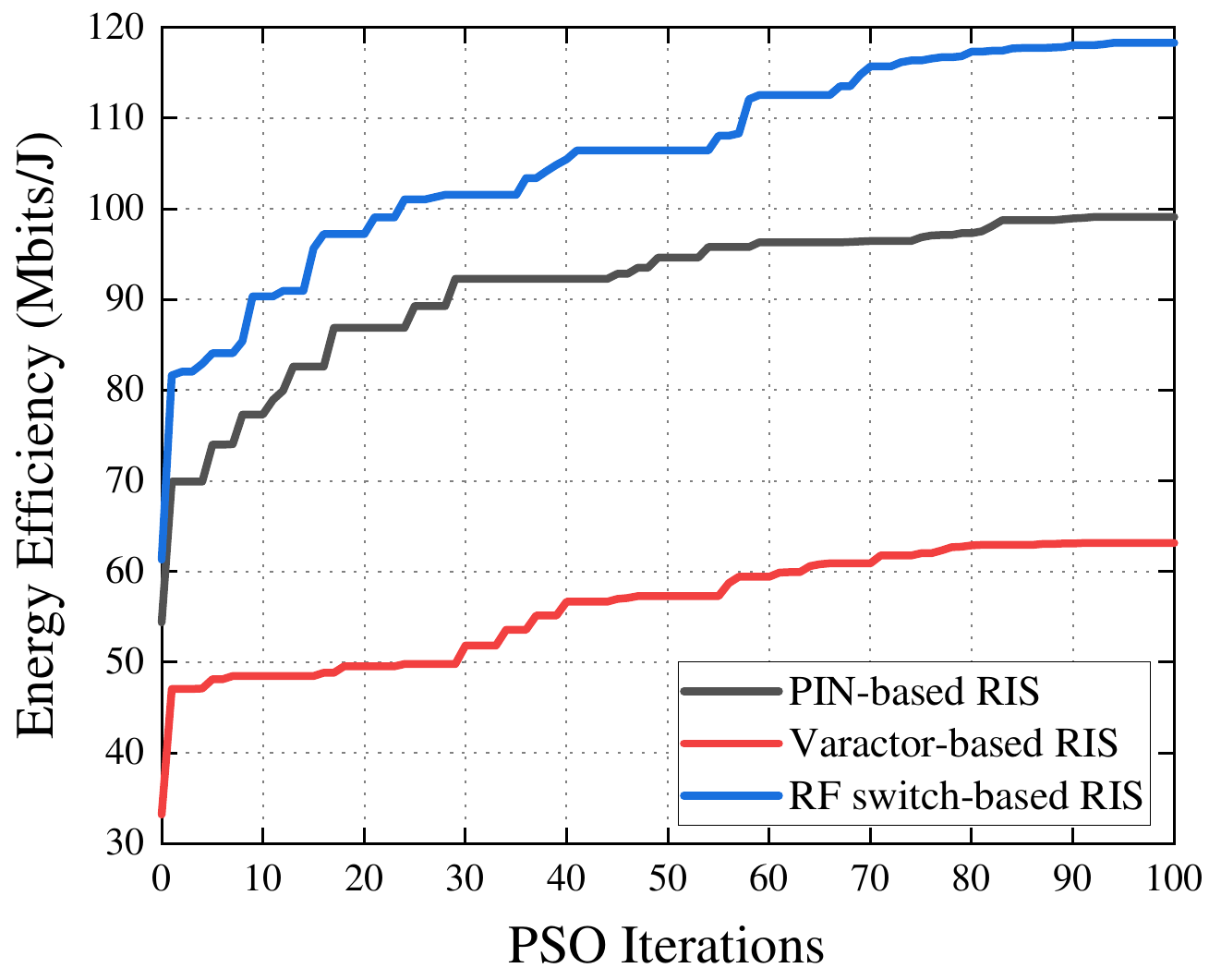}
    \vspace{-5pt}
    \caption{A group of convergence curves yielded by the proposed integer-based RIS near-field EE optimization algorithm, using PIN diodes, varactor diodes, and RF switches as the reconfigurable elements of the RIS.}
    \label{fig:PSO_Convergence}
\vspace{-10pt}
\end{figure}

\begin{table}[t]
\centering
\renewcommand{\arraystretch}{1.3}
\caption{Convergence parameters of integer-PSO}
\footnotesize
\label{table:PSO_parameter}
\vspace{-5pt}
\begin{tabular}{|l|c|c|c|c|c|} 
\hline
Parameters & $w$ & $c_1$ & $c_2$ & $d_1$ & $d_2$\\
\hline
Step 1 $\sim$ Step 65 & 0.6 & 1 & 1 & 0.6 & 0.6\\
Step 66 $\sim$ Step 85 & 0.4 & 0.9 & 1.1 & 0.4 & 0.4\\
Step 86 $\sim$ Step 100 & 0.2 & 0.8 & 1.2 & 0.1 & 0.1\\
\hline
\end{tabular}
\renewcommand{\arraystretch}{1.00}
\vspace{-10pt}
\end{table}

Fig.~\ref{fig:PSO_Convergence} presents the convergence curves for the proposed optimization algorithm adopting three different RIS reconfiguration methods when $M$=8, $N$=900, $K$=4, and $R_b$=3 bits. As they have the same RIS size and reconfiguration resolution in the same simulation scenario, their SEs are similar. Therefore, the differences in EEs between the different reconfiguration methods arise from their respective RIS power consumption models.
In this figure, the EE values at step~0 are based on the initial RIS configuration calculated by (\ref{equi:Initial_RIS_continuous}), with weights $r^{\rm{in}}_{1,m}$ and $r^{\rm{re}}_{1,k}$ set to be the same, ensuring that the needed phase compensations at RIS for each FBS antenna and UE are uniformly overlapped. After that, randomly generate these weights for (\ref{equi:Initial_RIS_continuous}) to give particles different initial RIS configurations to expand the hunting zone and avoid particles getting stuck in local optima. After performing precoding and power allocation at FBS for each updated RIS configuration, the maximal system EE among these initial 100 particles is set as the value at step~1, resulting in a sharp increment from step~0 to step~1.

From step~1 to step~100, 100 independent particles explore the solution space following the combined impact of individual optima and group optimum, adapting their RIS arrangements, FBS precoding, and power allocation in each step. The parameters in (\ref{equi:velocity}) and (\ref{equi:position}) to control the learning rate of particles between individual optimum and group optimum, along with the discard rates to expand exploration, are listed in Table~\ref{table:PSO_parameter}. The main purpose of setting these parameters is to facilitate effective exploration while ensuring all the particles will be converging to a common point by the end of iterations. Table~\ref{table:PSO_parameter} provides a valid parameter set, though other parameter sets may also be applicable. As the curves shown in Fig.~\ref{fig:PSO_Convergence}, our optimization method achieves an increment of approximately 82\% in EE for PIN-based RIS and around 90\% for the other two types of RISs.

\begin{figure}[t]
    \centering
    \includegraphics[width=0.8\columnwidth]{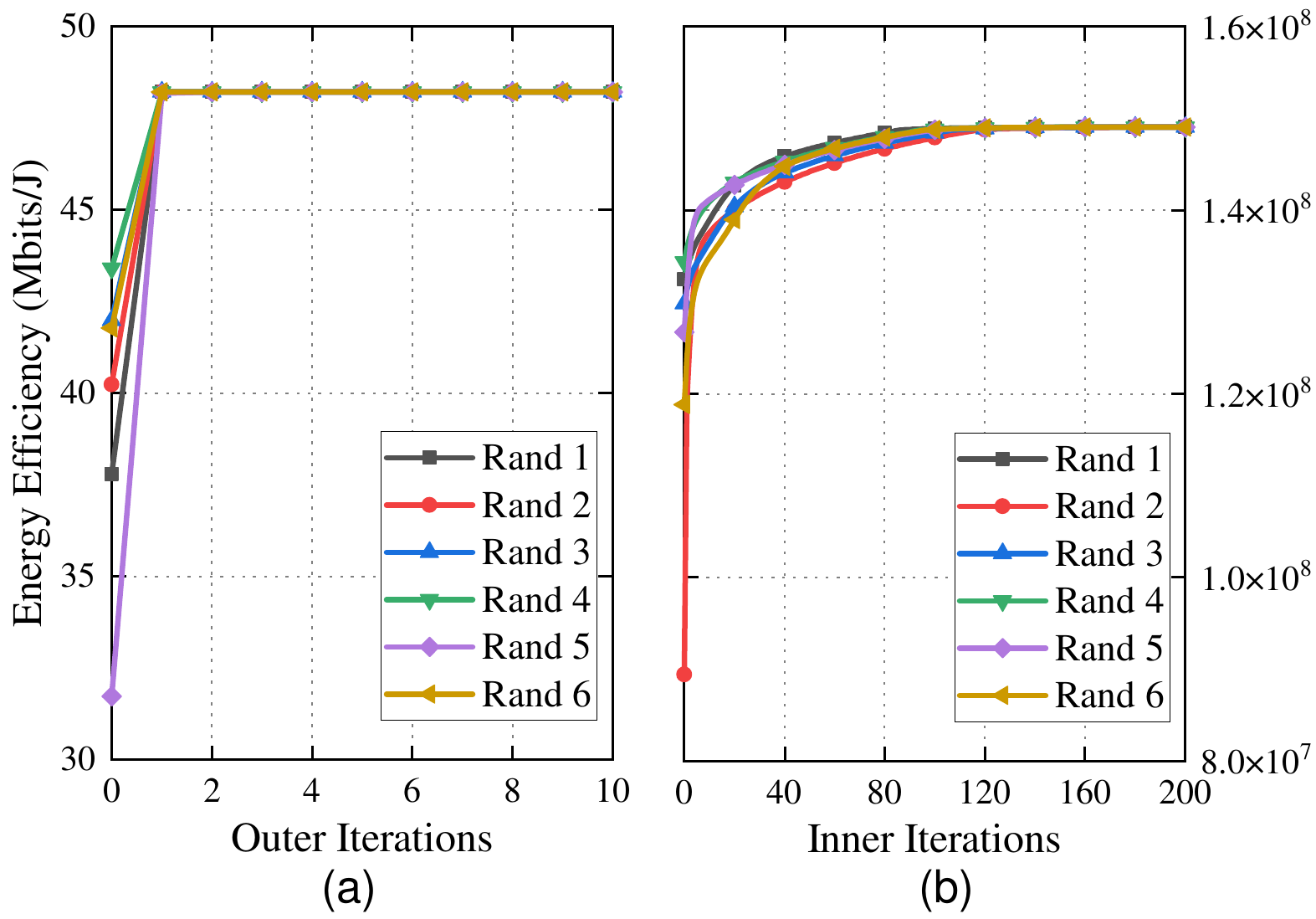}
    \vspace{-5pt}
    \caption{Convergence curves yielded by the proposed Dinkelbach-IQT algorithm, where (a) represents the outer iterations and (b) represents the inner iterations.}
    \label{fig:A1_Convergence}
    \vspace{-10pt}
\end{figure}

Fig.~\ref{fig:A1_Convergence} further illustrates the convergence performance of the proposed Dinkelbach-IQT algorithm, which is nested within the integer-PSO framework. In Fig.~\ref{fig:A1_Convergence}(a), the convergence of the outer iterations in Algorithm \ref{Algorithm 1} is evaluated under the same RIS configuration with six randomly initialized ${\boldsymbol {\rm p}}$, demonstrating rapid convergence within a limited number of iterations. Similarly, Fig.~\ref{fig:A1_Convergence}(b) presents the convergence of the inner iterations by setting $\eta_{\rm{EE}}=0$ in (\ref{PAQ1}). Despite differences in their initial values, the inner iterations consistently converge to the same local optimal solution. These results validate the convergence performance and effectiveness of the proposed Algorithm \ref{Algorithm 1}.

\subsubsection{Performance validation of proposed algorithm}
\begin{figure}[t]
    \centering
    \includegraphics[width=0.67\columnwidth]{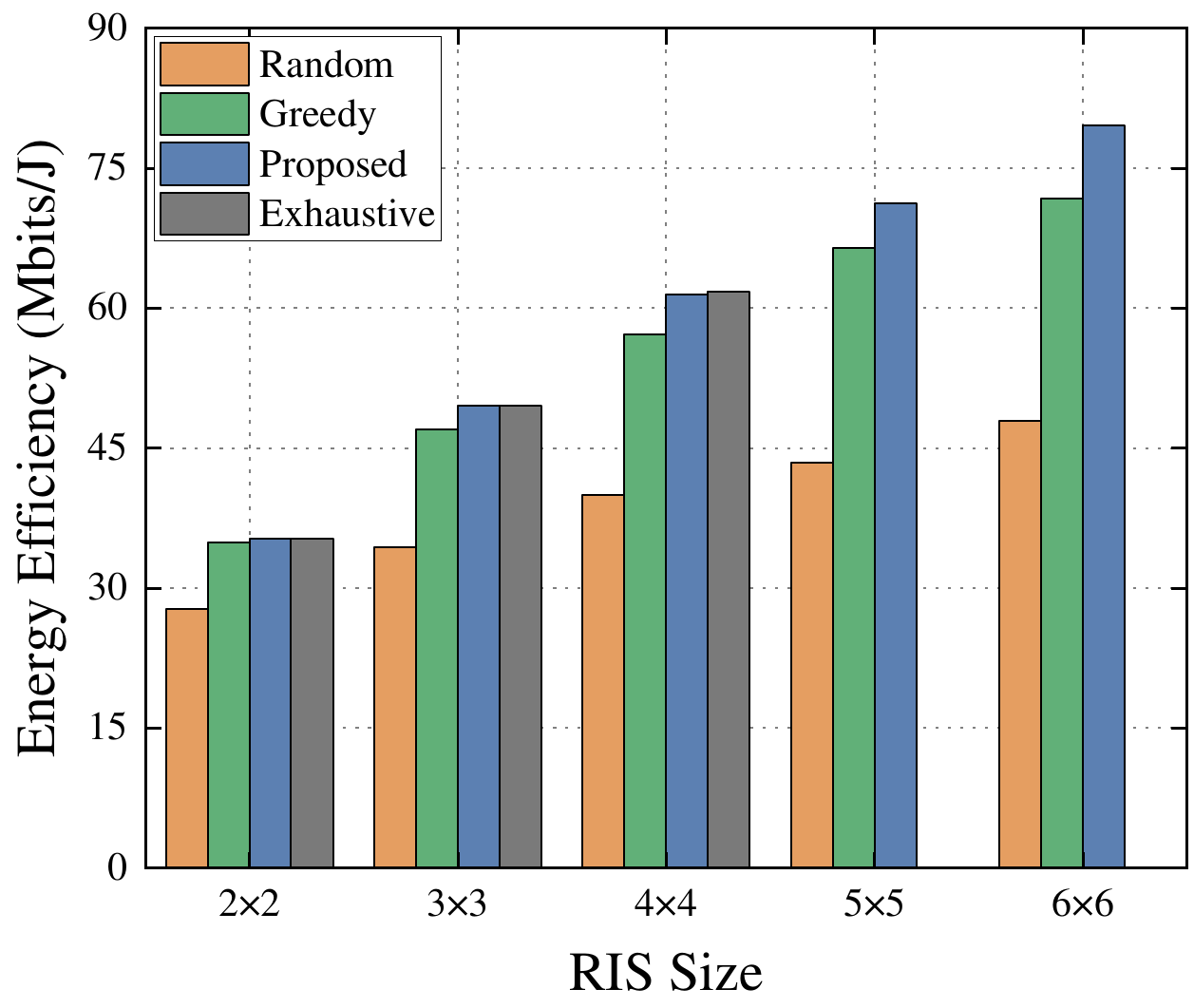}
    \vspace{-5pt}
    \caption{Average EE versus RIS size for small-scale validation of the proposed algorithm vs. random, greedy, and exhaustive search.}
    \label{fig:EE_Compare}
    \vspace{-10pt}
\end{figure}

To validate the optimization performance of the proposed algorithm, Fig.~\ref{fig:EE_Compare} presents a comparison with random search, greedy algorithm, and exhaustive search for RF switch-based RISs in optimizing phase-shift configurations with 1-bit resolution. Given that the computational time for exhaustive search becomes prohibitively long when the RIS size exceeds $4\times4$, only small-scale cases for RIS size from $2\times2$ to $6\times6$ are provided in this validation. As shown in the figure, the EE optimized by the proposed algorithm achieves near-optimal performance, closely approaching that of exhaustive search while significantly outperforming the random search and the greedy algorithm, validating the effectiveness and optimization performance of the proposed algorithm.

\subsubsection{Impact of transmit power}
\begin{figure}[t]
    \centering
    \includegraphics[width=0.7\columnwidth]{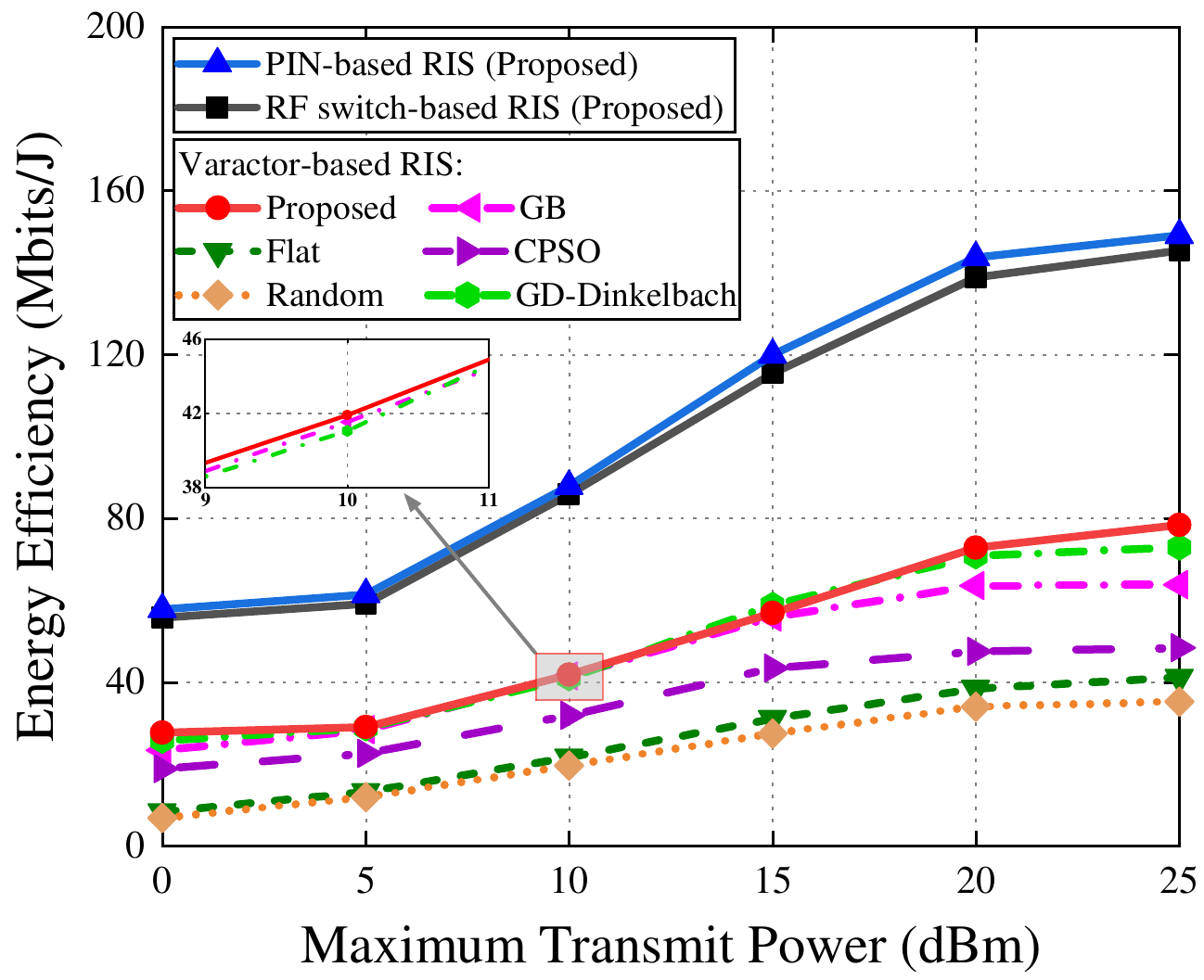}
    \vspace{-5pt}
    \caption{Average EE versus FBS transmit power for the proposed algorithm and three benchmarks, i.e., flat RIS, random RIS, and RIS optimized by gradient descent.}
    \label{fig:EE_TransmitPower}
    \vspace{-10pt}
\end{figure}

In Fig.~\ref{fig:EE_TransmitPower}, we depict the impact of the maximum transmit power of FBS on the system's EE when $M$=8, $N$=400, $K$=4, and $R_b$=1 bits. Besides, we further compare it with different benchmark schemes adopted in~\cite{9120336,shen2023multi,wang2021joint,9138463,huang2019reconfigurable, 9534477, 9066923, ma2024secure}:
\begin{itemize}[leftmargin=*]
    \item \textbf{Flat:} All RIS elements are fixed in the ``0" state, without applying any phase shifts.
    
    \item \textbf{Random:} RIS with randomly generated phase shifts.
    
    \item \textbf{GB:} RIS configurations optimized by the gradient-based (GB) algorithm, widely used for RIS reconfiguration~\cite{9534477, 9066923, 9120336}.

    \item \textbf{CPSO:} RIS configurations optimized by the continuous particle swarm optimization (CPSO) algorithm~\cite{ma2024secure}.

    \item \textbf{GD-Dinkelbach:} Alternative optimization where RIS configuration is optimized by gradient descent (GD) algorithm and BS power allocation is optimized by Dinkelbach algorithm~\cite{huang2019reconfigurable}.
\end{itemize}

Since the GB, GD, and CPSO algorithms are designed for continuous phase optimization in the literature, and varactor diode-based RISs exhibit negligible element power dissipation regardless of their real-time discrete configurations, we base our comparisons on varactor diode-based RISs. It is worth mentioning that while the Flat, Random, GB, and CPSO schemes differ in RIS reconfigurations, they all employ Algorithm~1 as a nested inner optimization for optimal BS power allocation, ensuring a fair comparison with the proposed integer-PSO algorithm. In comparison, the GD-Dinkelbach scheme follows a standard AO framework, using the customized GD algorithm for RIS reconfiguration and the Dinkelbach algorithm for power allocation, for comparison with the proposed nested AO framework. All results are based on discrete RIS phase shifts after quantization for a fair comparison.

As shown in Fig.~\ref{fig:EE_TransmitPower}, RISs based on PIN diodes or RF switches significantly outperform varactor-based RISs at 1-bit resolution with 400 RIS elements, due to the lower power consumption of their reconfigurable elements at low resolutions. As $P_{t}^{\max}$ increases from 0 dBm to 25 dBm, the EE of all these three RISs increases. This is because a higher $P_{t}^{\max}$ provides greater potential for FBS precoding and power allocation to enhance the desired signal and better control multi-user interference. Meanwhile, the increase in $P_{t}^{\max}$ is relatively small compared to other components of total power consumption within the given value range.

Compared to other benchmark schemes for varactor-based RISs, our proposed algorithm outperforms the Flat and Random schemes by around 90\% and 110\%, respectively, over the range of $P_{t}^{\max}$ from 10 to 15~dBm. Moreover, while GB and CPSO may achieve higher performance in a continuous phase-shift parameter space, they experience significant performance degradation due to quantization, particularly at low RIS resolutions. Therefore, by increasing the number of particles $N_{\rm p}$ and fine-tuning the convergence parameters for 1-bit resolution, the proposed integer-PSO algorithm significantly outperforms CPSO under the same computational complexity and slightly outperforms GB while requiring significantly lower computational complexity, with a complexity ratio of $\mathcal{O}(N_{\rm p} I_{\rm PSO})/\mathcal{O}(N I_{\rm GB})$=1/4, where $I_{\rm GB}$ is the number of iterations for the GB algorithm. When compared with GD-Dinkelbach, whose computational complexity $\mathcal{O}(I_{\rm alt}(I_{\rm GD}N^2+I_{\rm D}K))$~\cite{huang2019reconfigurable} exhibits quadratic growth with respect to the number of RIS elements $N$, with $I_{\rm alt}$, $I_{\rm GD}$, and $I_{\rm D}$ denoting the iteration numbers of AO, GD, and Dinkelbach, respectively, the proposed algorithm performs better in computation-constrained scenarios for large-scale RISs. As shown in Fig.~\ref{fig:EE_TransmitPower}, the proposed algorithm achieves comparable performance while maintaining significantly lower computational complexity.

\subsubsection{Impact of RIS reconfiguration resolutions}
\begin{figure}[t]
    \centering
    \includegraphics[width=0.7\columnwidth]{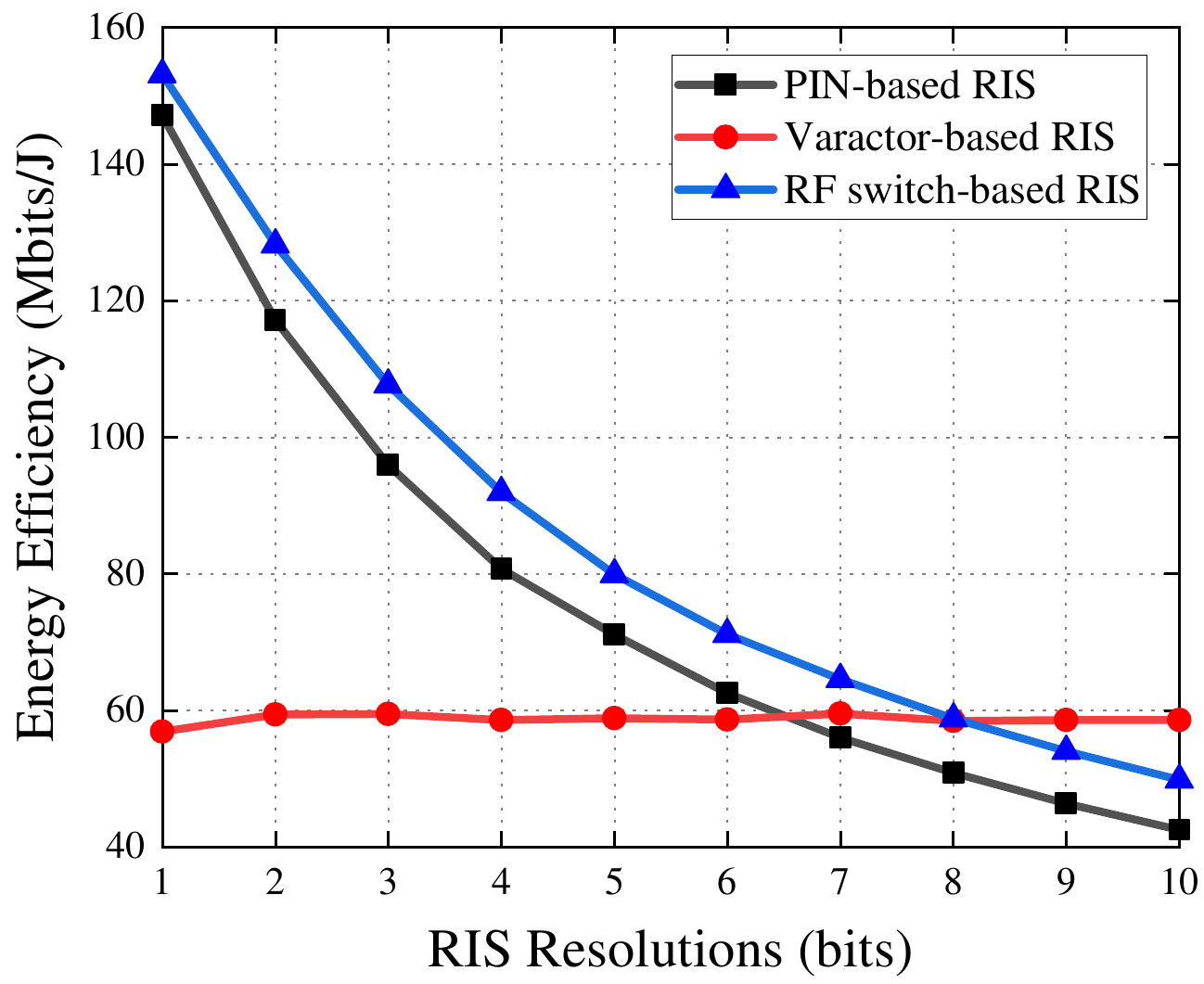}
    \vspace{-5pt}
    \caption{Average EE versus reconfiguration resolutions of reconfigurable elements.}
    \label{fig:EE_Resolution}
    \vspace{-10pt}
\end{figure}

Fig.~\ref{fig:EE_Resolution} shows the system's EE versus the resolution of reconfigurable elements when $M$=8, $N$=900, and $K$=4. Since the reflected phase shift of each reconfigurable element cannot be adjusted continuously but switches among discrete states, the available configurations and the performance of RIS-assisted communication systems are highly dependent on the resolution of the reconfigurable elements in bits.
While increasing the RIS resolution enhances the potential for yielding a higher SE within the same RIS size, it also increases the complexity of the reconfigurable element structure and the power required to turn the reconfigurable switches to work in the correct states. Therefore, as shown in Fig.~\ref{fig:EE_Resolution}, the EEs of PIN diodes-based RIS and RF switch-based RIS decrease by 70\% with the increment of reconfigurable element resolutions from 1-bit to 10-bit. This is because the increase in RIS resolution leads to a faster rise in power consumption compared to the improvement in SE.
Different from the other two reconfiguration methods, the EE of varactor diodes-based RIS slightly improves from 57 Mbits/J to 59 Mbits/J when the resolution increases from 1-bit to 2-bit and remains stable when further increasing the resolution. This is because the power consumption of varactor diodes-based RIS is mostly by its energy-greedy driving circuits, with little impact from the mounted varactor diodes operating RIS configurations. 
Comparing these three different reconfiguration methods, the PIN diodes-based RIS and RF switch-based RIS have a much higher EE when RISs have low resolutions from 1-bit to 6-bit since their power consumption is lower than varactor diodes-based RIS despite they can achieve a similar SE.
In contrast, varactor diodes-based RIS outperforms with high resolutions, beating the other two types of RIS when the reconfiguration resolution is higher than 8-bit under our parameter setup.
However, high resolutions for RISs may not be realistic in real-world communication applications because of the poor EE.

\begin{figure*}[t]
\centering 
    \subfloat[\label{fig:EE_Number_1bit}]{
      \begin{minipage}[t]{0.3\linewidth}
        \includegraphics[width=0.9\columnwidth]{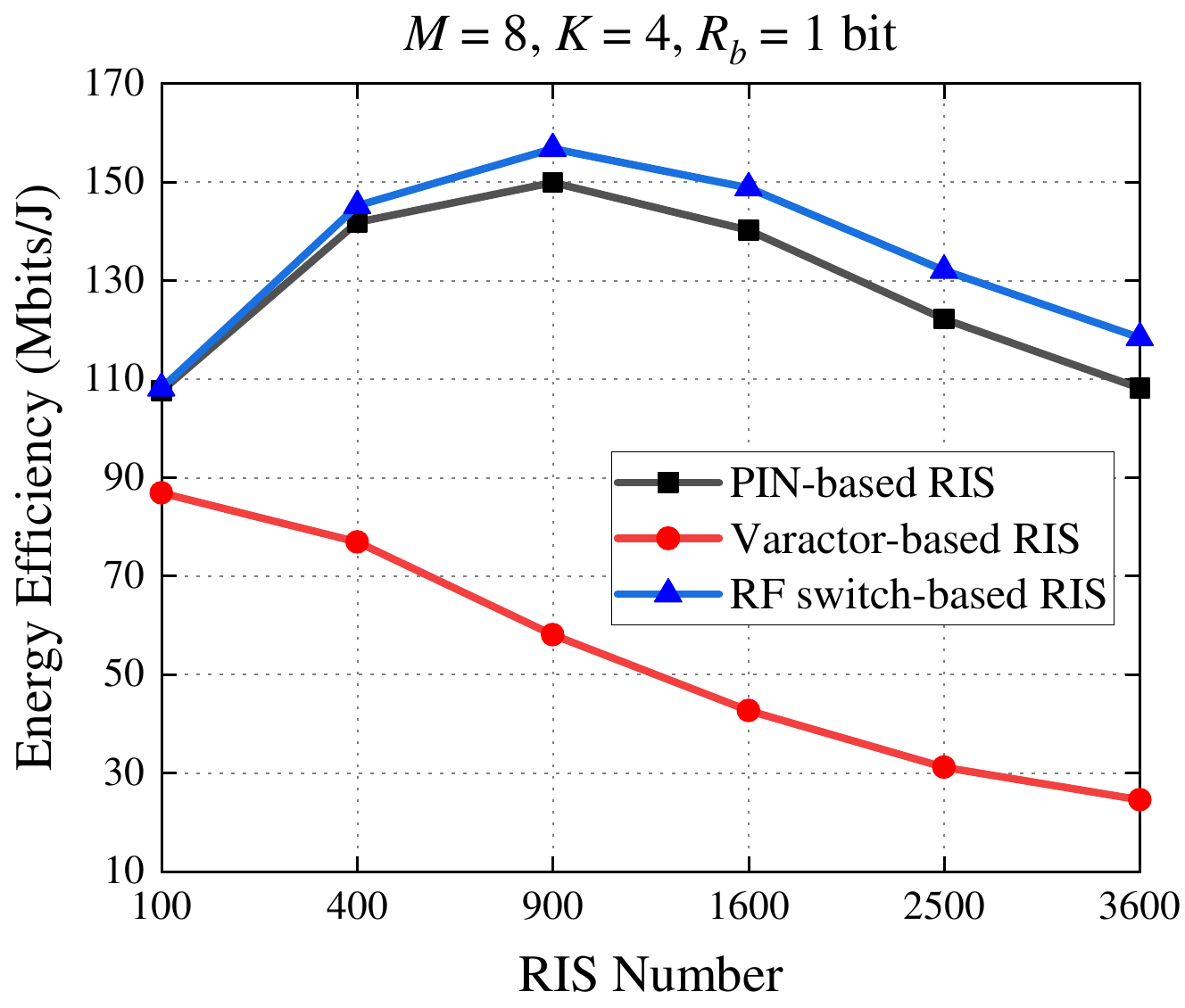}   
      \end{minipage}%
      }
    \subfloat[\label{fig:EE_Number_2bit}]{
      \begin{minipage}[t]{0.3\linewidth}   
        \includegraphics[width=0.9\columnwidth]{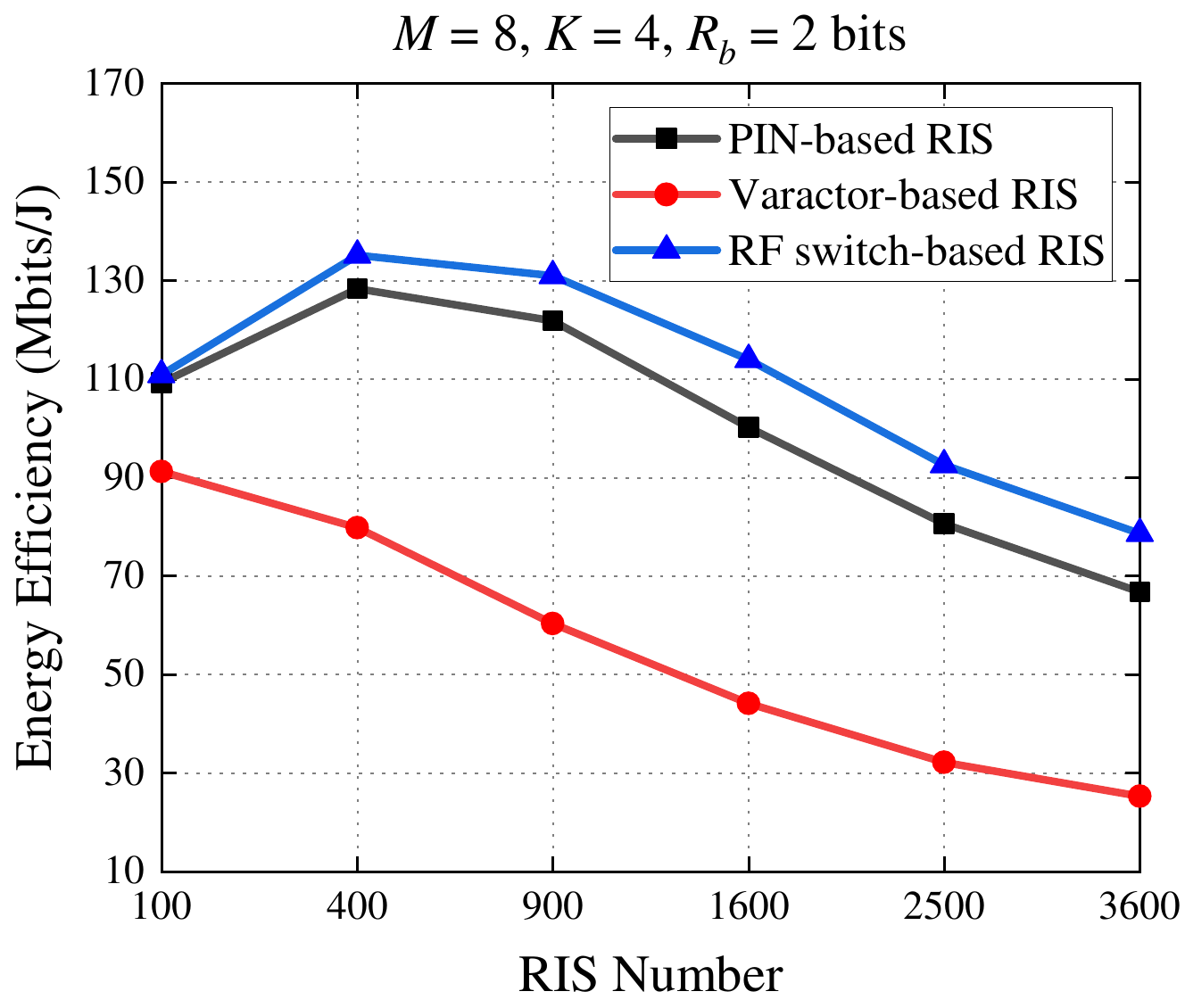}   
      \end{minipage} 
      }
    \subfloat[\label{fig:EE_Number_3bit}]{
      \begin{minipage}[t]{0.3\linewidth}   
        \includegraphics[width=0.9\columnwidth]{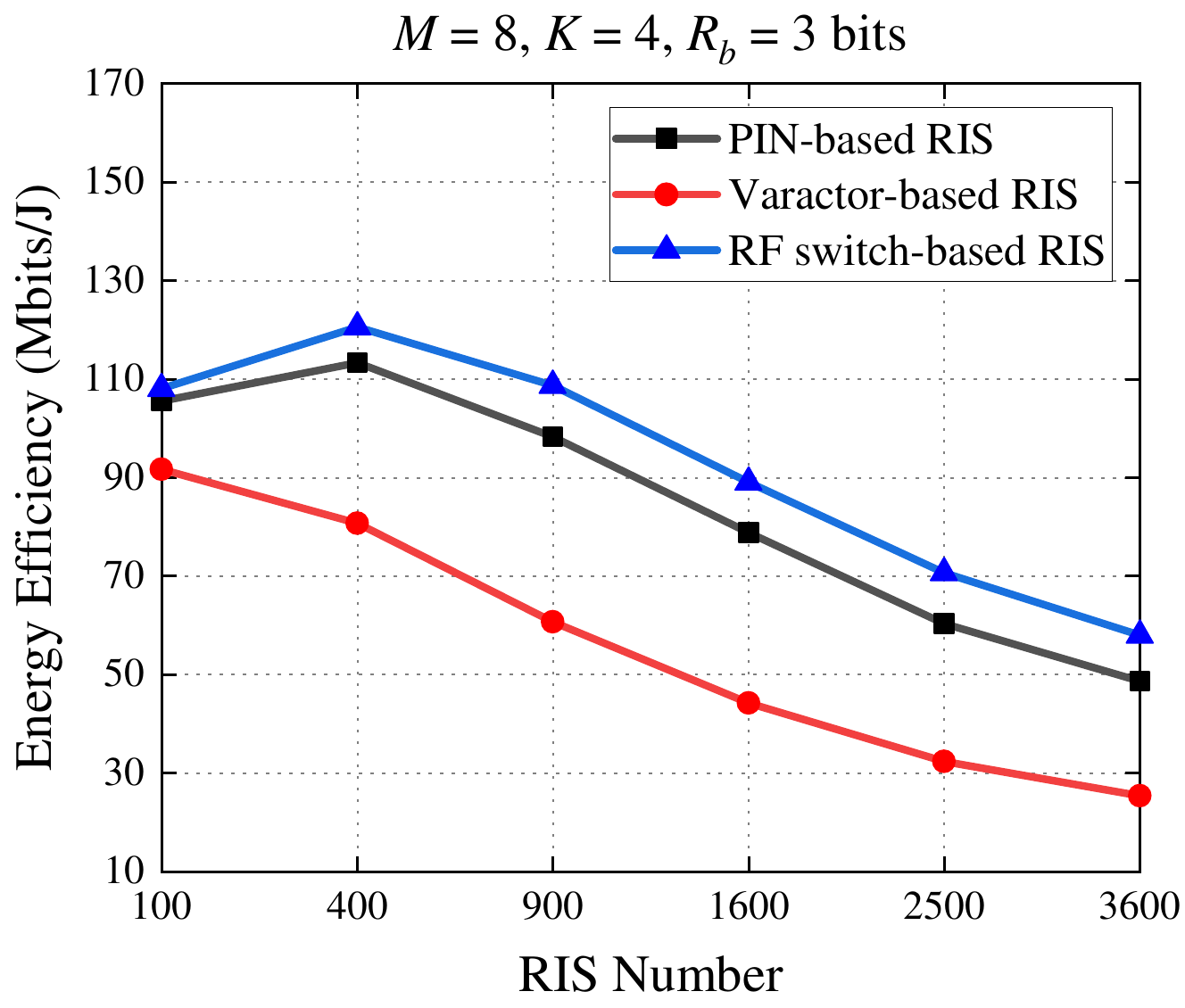}   
      \end{minipage} 
      }
      \vspace{-5pt}
    \subfloat[\label{fig:EE_Number_4bit}]{
      \begin{minipage}[t]
      {0.3\linewidth}
        \includegraphics[width=0.9\columnwidth]{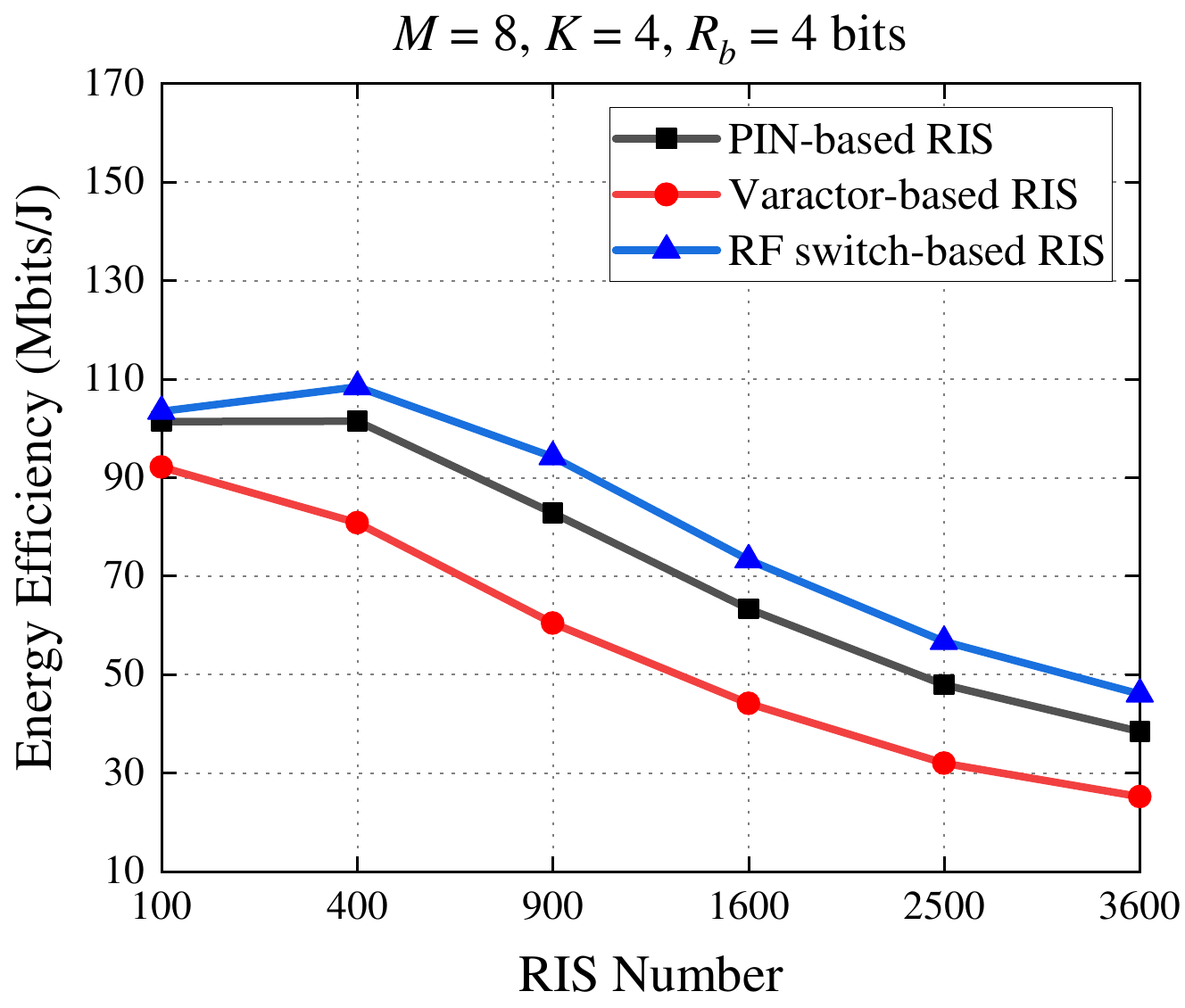}   
      \end{minipage}%
      }
    \subfloat[\label{fig:EE_Number_5bit}]{
      \begin{minipage}[t]{0.3\linewidth}   
        \includegraphics[width=0.9\columnwidth]{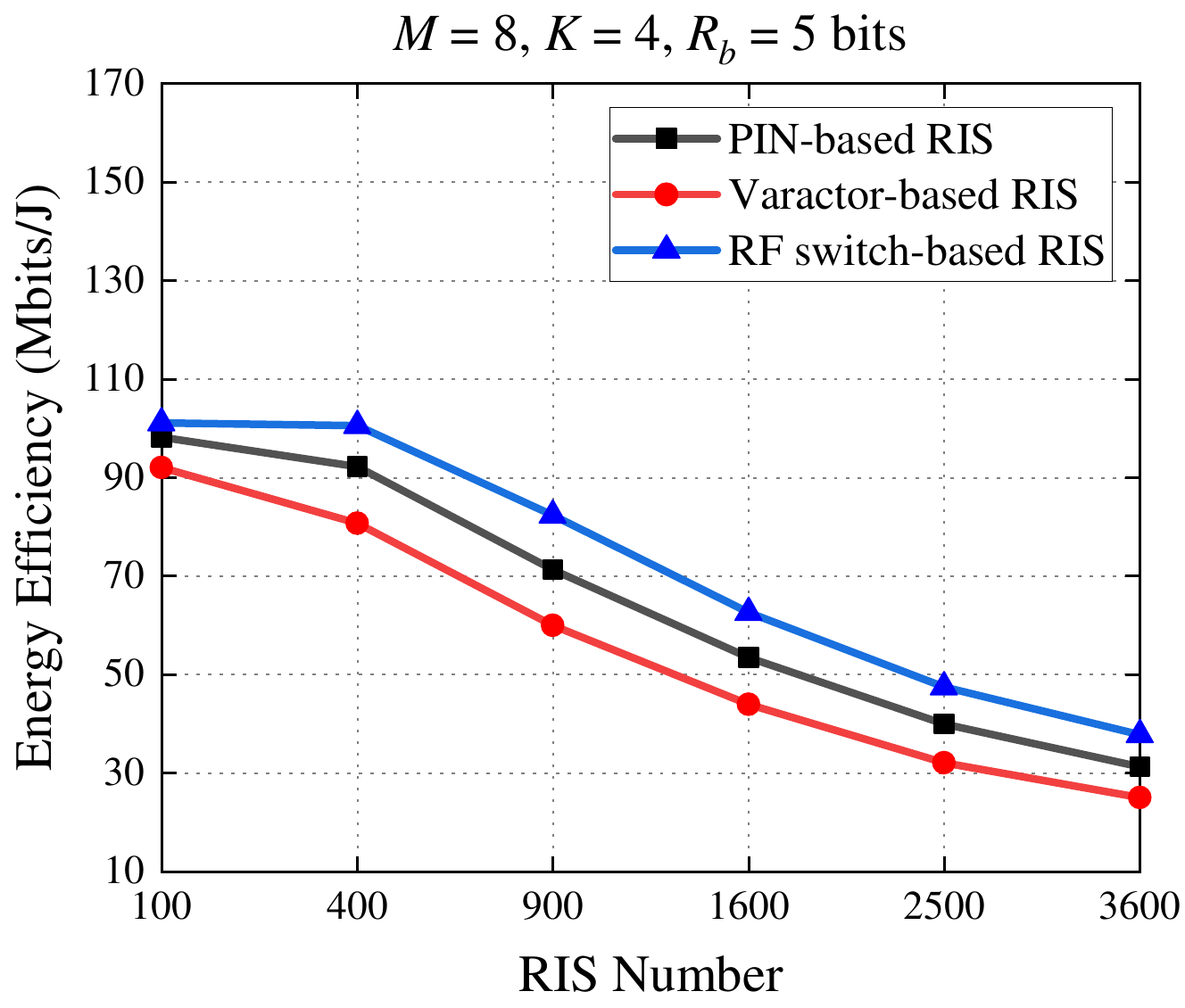}   
      \end{minipage} 
      }
    \subfloat[\label{fig:EE_Number_6bit}]{
      \begin{minipage}[t]{0.3\linewidth}   
        \includegraphics[width=0.9\columnwidth]{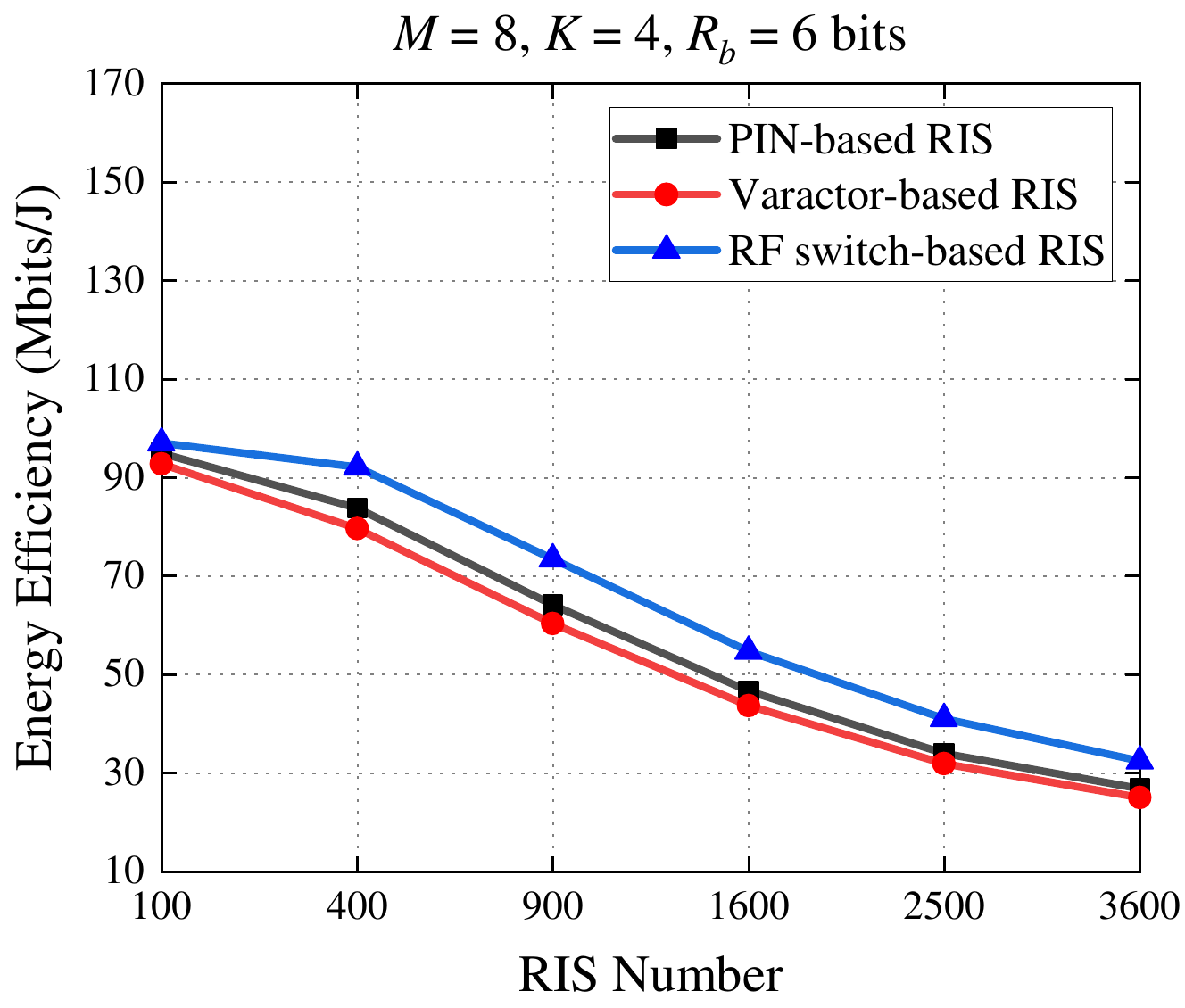}   
      \end{minipage} 
      }
\caption{Average EE versus RIS number under (a) 1-bit, (b) 2-bit, (c) 3-bit, (d) 4-bit, (e) 5-bit, and (f) 6-bit reconfiguration resolutions.} \label{fig:EE_Number}
\vspace{-10pt}
\end{figure*}

\subsubsection{Impact of RIS sizes}
Fig.~\ref{fig:EE_Number} illustrates the impact of both RIS size and resolution. From Fig.~\ref{fig:EE_Number}(a) to Fig.~\ref{fig:EE_Number}(f), the EEs versus reconfigurable element numbers under different resolutions are presented. In these sub-figures, the overall performance of PIN-based RIS and RF switch-based RIS degrades and approaches varactor-based RIS when the resolution increases from 1-bit to 6-bit, further verifying the observation made from Fig.~\ref{fig:EE_Resolution}.
When concentrating on the cases with low resolutions, the RIS size displays a significant impact on the system EE at a fixed resolution. The 1-bit resolution gives reconfigurable elements only two discrete phase states for reconfiguration, which limits the potential of the entire RIS array. However, 1-bit resolution also results in the lowest power dissipated by reconfigurations for the RIS board. Under these circumstances, even though expanding the RIS size introduces a square-increased element number to consume much higher power, the system EE of PIN-based RIS and RF switch-based RIS still improves from 108~Mbits/J to around 150~Mbits/J when the RIS size is expanded from $10\times 10$ to $30\times 30$. As shown in Fig.~\ref{fig:EE_Number}(a), the number of elements between 400 and 2,500 positively affects the EE of 1-bit RIS, with RISs having approximately 1,000 elements capable of achieving the highest EE.

With the resolution increases from 1 bit to 2 bits as shown in Fig.~\ref{fig:EE_Number}(b), PIN-based and RF switch-based RISs show enhanced EEs with a low RIS size of 100 elements. This is because the 2-bit resolution provides each reconfigurable element with two additional discrete phase states for better phase compensation on RISs, bringing about a more significant increase in SE compared to power consumption.
However, when it comes to larger RIS sizes, this advantageous feature disappears as the achieved increment in SE cannot compensate for the elevated RIS power dissipation. Therefore, even though increasing the element number from 100 to 400 can still provide better EE for 2-bit RIS, the 2-bit RIS cannot outperform the 1-bit RIS in terms of EE when the element number exceeds 400.
Along with further increased resolution from Fig.~\ref{fig:EE_Number}(c) to Fig.~\ref{fig:EE_Number}(f), the EE of PIN-based and RF switch-based RISs continues to decrease. As a result, the expansion of RIS size brings about more energy consumption to the system rather than the EE improvement.
 
For varactor diode-based RISs, the power dissipated by their elements to achieve reconfigurations is so small that it can be neglected, resulting in constant system power consumption regardless of the specific RIS resolutions. Since the increase in resolution benefits SE, varactor diode-based RISs with more than 2-bit resolution achieve better EE with a low RIS size, such as with 100, 400, and 900 elements as shown in Fig.~\ref{fig:EE_Number}. However, in contrast to the increase in resolution, the expansion of RIS size necessitates more driving circuits, leading to a more significant reduction in the EE of varactor diode-based RISs compared to the other two reconfiguration methods with the increment of RIS size.

\subsection{Evaluation and Discussion}
From the aforementioned simulation results, we can make two key observations that may aid in the design of real-world RIS-assisted near-field communication systems.

First, although RF switches-based RISs and PIN diodes-based RISs have different power consumption characteristics, where the power dissipation of the former is unrelated to real-time RIS configurations, while the latter is real-time configuration-dependent, they exhibit almost consistent EE performance. The gaps between these two reconfiguration methods in the figures above are due to the specific power consumption parameters used in our work, and these differences may be eliminated with other parameter setups.
Compared to RF switch-based and PIN diode-based RISs, varactor diode-based RISs have much higher power consumption when the reconfiguration resolution is low. However, their power dissipation does not vary with specific RIS resolutions, resulting in more stable power consumption across different resolutions. As the EEs of RF switch-based and PIN diode-based RISs decrease with increasing resolutions due to elevated power consumption, varactor diode-based RISs can finally surpass them in both power consumption and EE when the reconfiguration resolution goes high.

Second, the element number and resolution significantly influence the performance of RIS-assisted near-field communications since increasing them can enhance the maximum achievable SEs. However, this comes at the cost of higher energy consumption, which may lead to a decrease in overall system EE even if SE improves. Therefore, finding the equilibrium point for the highest EE is important. Based on our simulation setup and results, if there are no limitations on the deployment size of RIS, 1-bit RF switch-based or PIN diode-based RISs with around 1,000 reconfigurable elements could be the most suited paradigms for indoor near-field communication scenarios. If the deployment area is compact, a smaller RIS size with around 100 reconfigurable elements using 2-bit or 3-bit resolutions is recommended for each of these three reconfiguration methods. In addition, appropriately adjusting the maximum transmit power at BSs can also benefit the system's EE.

\section{Conclusions}
\label{sec:conclusions}
In this work, we studied the EE of indoor RIS-assisted multiuser downlink near-field wireless communication systems. To better align with the real-world power characteristics of RISs, we adopted three more practical power consumption models of RIS that comprehensively considers various factors, including the number and resolution of reconfigurable elements, RIS reconfiguration methods, driving circuits, RIS controller, and the maximum transmit power of BSs. This is the first work to analyze the system performance under such a practical power consumption models for various reconfiguration methods. Furthermore, we optimized the RIS phase configurations in the discrete parameter space and the BS power allocation in the continuous parameter space. Our proposed integer-based RIS configuration optimization can achieve a higher EE compared to benchmark schemes operating in the continuous parameter space when considering the discrete phases of RIS in real-world applications.
By comparing the EE performance of RISs with different element numbers, resolutions, and reconfiguration methods, we evaluated the balance between SE and power consumption, identifying the optimal RIS architecture setups with the highest EE for various deployment requirements. Additionally, we systematically discussed the power consumption characteristics and system EEs for RISs using PIN diodes, varactor diodes, and RF switches, observing their significantly different power dissipation despite having similar parameters.

{\appendix[Proof of Lemma 1]
We first formulate the Lagrangian function of problem $\mathcal{P}6$ as follows
\begin{equation}\label{A1}
L\left( {\boldsymbol {\rm p}}, \rho \right) = G\left( {\boldsymbol {\rm p}},\mathbf{\Gamma }^{[t+1]},\mathbf{y}^{[t+1]} \right) +\rho \left( P_{t}^{\max}-\sum_{k=1}^K{P_k} \right),
\end{equation}
where $\rho$ denotes a Lagrangian multiplier. The partial derivative of $L\left( {\boldsymbol {\rm p}}, \rho \right)$ with respect to $P_k$ is obtained by
\begin{align}\label{A2}
\frac{\partial L\left( {\boldsymbol {\rm p}}, \rho \right)}{\partial P_k}= &\frac{y_{k}^{[t+1]}\sqrt{\vartheta ( 1+\Gamma _{k}^{[t+1]} ) \zeta _{k,k}}}{\sqrt{P_k}}
\notag \\
&-\sum_{j=1}^K{\left( y_{j}^{[t+1]} \right) ^2\zeta _{j,k}}-\eta _{\rm{EE}}^{[n]}\xi -\mu.
\end{align}

Thus, solving $\partial L\left( {\boldsymbol {\rm p}}, \rho \right)/\partial P_k = 0$ yields optimal power allocation results with a parameter $\rho$ as shown in (\ref{PA13}). Evidently, $P_{k}^{[t+1]}(\rho)$ is an inverse quadratic function of $\rho$. According to the following Karush–Kuhn–Tucker (KKT) conditions
\begin{equation}\label{A3}
\begin{cases}
	\sum_{k=1}^K{P_{k}^{[t+1]}\left( \rho \right)}\le P_{t}^{\max},\\
	\rho \left( P_{t}^{\max}-\sum_{k=1}^K{P_{k}^{[t+1]}\left( \rho \right)} \right) =0,\\
	\rho \ge 0,\\
\end{cases}
\end{equation}
the Lagrangian multiplier $\rho$ can be updated according to the rule given in (\ref{PA14}), which completes the proof of \textit{Lemma 1}.
}
 
\bibliographystyle{IEEEtran}
\bibliography{IEEEabrv,references}


 





\end{document}